\documentclass[final,3p,times]{elsarticle}

\usepackage{float}
\usepackage{graphicx}
\usepackage[bookmarksnumbered=true,
bookmarksopen=true,
colorlinks=true,
citecolor=blue,
linkcolor=red,
anchorcolor=green,
urlcolor=blue]{hyperref}
\usepackage{amssymb,amsmath}
\usepackage{color}
\usepackage{threeparttable}
\usepackage{multirow}
\usepackage{booktabs}

\newcommand{\tabincell}[2]{\begin{tabular}{@{}#1@{}}#2\end{tabular}}

\journal{\url{arXiv.org}}
\begin{document}

\begin{frontmatter}

\title{\emph{Linked Ego Networks}: Improving Estimate Reliability and Validity with Respondent-driven Sampling}

\author[kiphs,susocial,nudt]{Xin Lu\corref{cor1}}
\ead{lu.xin@sociology.su.se}

\cortext[cor1]{Address for correspondence: Xin Lu, Department of Sociology, Stockholm University, SE-106 91, Stockholm, Sweden.}
\address[kiphs]{Department of Public Health Sciences, Karolinska Institutet, Stockholm, Sweden}
\address[susocial]{Department of Sociology, Stockholm University, Stockholm, Sweden}
\address[nudt]{College of Information Systems and Management, National University of Defense Technology, Changsha, China}

\begin{abstract}

Respondent-driven sampling (RDS) is currently widely used for the study of HIV/AIDS-related high risk populations. However, recent studies have shown that traditional RDS methods are likely to generate large variances and may be severely biased since the assumptions behind RDS are seldom fully met in real life. To improve estimation in RDS studies, we propose a new method to generate estimates with ego network data, which is collected by asking RDS respondents about the composition of their personal networks, such as ``\emph{what proportion of your friends are married?}''. By simulations on an extracted real-world social network of gay men as well as on artificial networks with varying structural properties, we show that the new estimator, $RDSI^{ego}$ shows superior performance over traditional RDS estimators. Importantly, $RDSI^{ego}$ exhibits strong robustness to the preference of peer recruitment and variations in network structural properties, such as homophily, activity ratio, and community structure. While the biases of traditional RDS estimators can sometimes be as large as $10\%\sim20\%$, biases of all $RDSI^{ego}$ estimates are well restrained to be less than $2\%$. The positive results henceforth encourage researchers to collect ego network data for variables of interests by RDS, for both hard-to-access populations and general populations when random sampling is not applicable. The limitation of $RDSI^{ego}$ is evaluated by simulating RDS assuming different level of reporting error.

\end{abstract}

%
%

\end{frontmatter}

\section{Introduction}

In many forms of research, there is no list of all members for the studied population (i.e., a \emph{sampling frame}) from which a random sample may be drawn and estimates about the population characteristics may be inferred based on the select probabilities of sample units. Non-probability sampling methods may be used for for such situations, such as key informant sampling~\cite{ref11Deaux1985}, targeted/location sampling~\cite{ref12Watters1989}, and snowball sampling~\cite{ref13Erickson1975}. However, these methods all introduce a considerable selection bias, which impairs generalization of the findings from the sample to the studied population~\cite{ref14Heckathorn1997, ref15Magnani2005}. Respondent-driven sampling (RDS) is an alternative method that is currently being used extensively in public health research for the study of \emph{hard-to-access} populations, e.g., injecting drug users (IDUs), men who have sex with men (MSM) and sex workers (SWs). With a link-tracing network sampling design, the RDS method provides unbiased population estimates as well as a feasible implementation, making it the state-of-the-art sampling method for studying \emph{hard-to-access} populations~\cite{ref1Johnston2008,ref2Wejnert,ref3Lansky2007,ref4Kogan2011,ref5Wejnert2008}.


RDS starts with a number of pre-selected respondents who serve as ``seeds''. After an interview, the seeds are asked to distribute a certain number of coupons (usually 3) to their friends who are also within the studied population. Individuals with a valid coupon can then participate in the study and are provided the same number of coupons to distribute. The above recruitment process is repeated until the desired sample size is reached~\cite{ref14Heckathorn1997}. In a typical RDS, information about who recruits whom and the respondents' number of friends within the population (degree)
are also recorded for the purpose of generating population estimates from the sample~\cite{ref16Heckathorn2002,ref17Salganik2004}.

Suppose a RDS study is conducted on a connected network with the additional assumptions that (i) network links are undirected, (ii) sampling of peer recruitment is done with replacement, (iii) each participant recruits one peer from his/her neighbors, and (iv) the peer recruitment is a random selection among all the participant's neighbors. Then the RDS process can be modeled as a Markov process, and the composition of the sample will stabilize and be independent of the properties of the seeds~\cite{ref17Salganik2004,ref18Heckathorn2007,ref19Volz2008}. Following this, the probability for each node to be included in the RDS sample is proportional to its degree. Specifically, for a given sample $U = \{ {v_1},{v_2}, \ldots ,{v_n}\} $, with ${{n}_{A}}$ being the number of respondents in the sample with property $A$ (e.g., HIV-positive) and ${n_B} = n - {n_A}$ being the rest. Let $\{ {d_1},{d_2}, \ldots ,{d_n}\} $ be the respondents' degree and $S=\left[ \begin{matrix}
   {{s}_{AA}} & {{s}_{AB}}  \\
   {{s}_{BA}} & {{s}_{BB}}  \\
\end{matrix} \right]$ be the recruitment matrix observed from the sample, where ${s_{XY}}$ is the proportion of recruitments from group $X$ to group $Y$ (for the purpose of this paper, we consider a binary property such that each individual belongs either to group $A$ or $B$). Then the proportion of individuals belonging to group $A$ in the population, $P_A^*$, can be estimated by~\cite{ref17Salganik2004,ref19Volz2008}:

\begin{equation}
{\hat P_A} = \frac{{{s_{BA}}{{\hat{\bar D}}_B}}}{{{s_{AB}}{{\hat {\bar D}}_A} + {s_{BA}}{{\hat{\bar D}}_B}}}{\text{ }{(RDSI),}}\label{eq1}
\end{equation}

or

\begin{equation}
{\hat P_A} = \frac{{\sum\limits_{{v_i} \in A \cap U} {d_i^{ - 1}} }}{{\sum\limits_{{v_i} \in U} {d_i^{ - 1}} }}{\text{ }{(RDSII),}}\label{eq2}
\end{equation}

where ${{\hat{\bar{D}}}_{A}}=\frac{{{n}_{A}}}{\sum\limits_{{{v}_{i}}\in A\bigcap U}{d_{i}^{-1}}}$ and ${\hat{\bar D}_B} = \frac{{{n_B}}}{{\sum\limits_{{v_i} \in B \cap U} {d_i^{ - 1}} }}$ are the estimated average degrees for individuals of group $A$ and $B$ in the population. Both estimators give asymptotically unbiased estimates. The estimation procedure above is also called a re-weighted random walk (RWRW) in other fields~\cite{ref20Gjoka}.

The methodology of RDS is nicely designed; however, the assumptions underlying the RDS estimators are rarely met in practice~\cite{ref2Wejnert,ref8Tomas2011,ref10Goel2010}. For example, empirical RDS studies use more than one coupon and sampling is conducted without replacement, that is, each respondent is only allowed to participate once. A comprehensive evaluation has been made by Lu et al~\cite{ref7Lu2012}, where the effects of violation of assumptions (i)$\sim$(iv), as well as the effect of selection and number of seeds and coupons, were evaluated one by one, by simulated RDS process on an empirical MSM network as well as artificial networks with known population properties. They have shown that when the sample size is relatively small ($<10\%$ of the population), RDS estimators have a strong resistance to violations of certain assumptions, such as low response rate and errors in self-reporting of degrees, and the like. On the other hand, large bias and variance may result from differential recruitments, or from networks with irreciprocal relationships. When the sample size is relatively large ($>50\%$ of the population), similar results were also found by Gile and Handcock~\cite{ref6Gile2010}, where they focused on the sensitivity of RDS estimators to the selection of seeds, respondent behavior and violation of assumption (ii).


It was not until recently that researchers found the variance in RDS may have been severely underestimated~\cite{ref21Salganik2006}. In a study by Goel and Salganik~\cite{ref10Goel2010} based on simulated RDS samples on empirical networks, they found that the RDS estimator typically generates five to ten times greater variance than simple random sampling~\cite{ref21Salganik2006}. Moreover, McCreesh et al~\cite{ref9McCreesh2012} conducted a RDS study on male household heads in rural Uganda where the true population data was known, and they found that only one-third of RDS estimates outperformed the raw proportions in the RDS sample, and only 50\%-74\% of RDS 95\% confidence intervals, calculated based on a bootstrap approach for RDS, included the true population proportion.

For the above reasons, there has been an increasing interest in developing new RDS estimators to improve the performance of RDS. For example, Gile~\cite{ref22Gile2011} developed a successive-sampling-based estimator for RDS to adjust the assumption of sampling with replacement and demonstrated its superior performance when the size of the population is known. Lu et al~\cite{ref23Lu2012} proposed a series of new estimators for RDS on directed networks, with known indegree difference between estimated groups. Both of the above estimators can be used as a sensitivity test when the required population parameters are not known.

Both the traditional $RDSI$, $RDSII$ estimators, and the estimators newly developed by Gile et al~\cite{ref22Gile2011,ref24Krista2011} and Lu et al~\cite{ref23Lu2012} utilize the same information collected by standard RDS practice, that is, the recruitment matrix $S$, respondents' degree, and sample property. There is however scope to improve estimates dramatically if data on the composition of respondents' ego networks can be put to use. Such data has already been collected for other purposes in many RDS studies. For example, in a RDS study of MSM in Campinas City, Brazil, by de Mello et al~\cite{ref25deMello2008}, respondents were asked to describe the percentage of certain characteristics among their friends/acquaintances, such as disclosure of sexual orientation to family, HIV status, and the like. In a RDS study of opiate users in Yunnan, China, various information about supporting, drug using, and sexual behaviors between respondents and their network members were collected~\cite{ref26Li2011}. One of the most thorough RDS studies utilizing ego network information was done by Rudolph et al~\cite{ref27Rudolph2011}, in which they asked the respondents to provide extensive characteristics for each alter within their personal networks such as demographic characteristics, history of incarceration, and drug injection and crack and heroin use.

Aiming to improve the RDS estimator, we will focus on how to integrate this additional information in the estimation process to generate improved population estimates. The rest of this paper is organized as follows. In Section 2, we develop a new estimator that integrates traditional RDS data with egocentric data; in Section 3, we describe network data used for simulation and study design; in Section 4, we evaluate the performance of the new estimator by simulated RDS processes under various settings; and in Section 5, we summarize and draw our conclusions.

\section{$RDSI^{ego}$: estimator for RDS with egocentric data}

\begin{figure}[ht]
 \begin{center}
 \centerline{\includegraphics[width=1\textwidth]{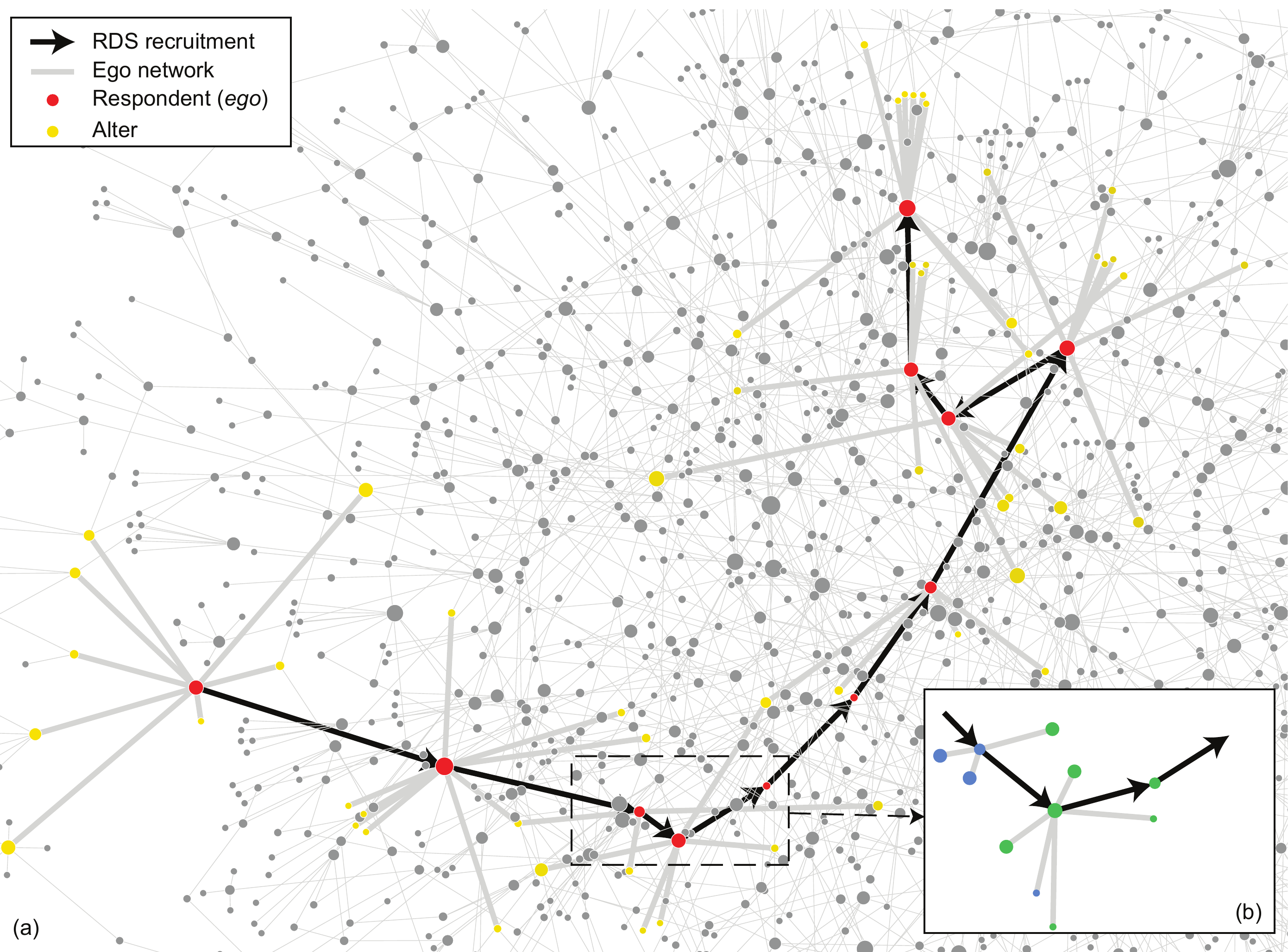}}
 \caption{A RDS chain with egocentric data. (a) RDS on a network. Red nodes are those that participated in the RDS survey, and yellow nodes are ego network composition inferred by participants; (b) A partial RDS chain with color representing properties of nodes.}\label{fig1}
 \end{center}
 \end{figure}

The ego networks from a RDS sample differ from general egocentric data collected in many sociological surveys~\cite{ref28Tom2012} in the way that each ``ego'' is connected with (recruited by) its recruiter. For example, in a partial chain of RDS as illustrated in \autoref{fig1}, participants ${v_i}$, ${v_j}$, ${v_k}$, are asked to provide personal network compositions and ${v_j}$ and $v_k$ are recruited by ${v_i}$, ${{v}_{j}}$, respectively.

For each respondent ${{v}_{i}}$ in a RDS sample $U = \{ {v_1},{v_2}, \ldots ,{v_n}\} $, let $n_i^A$, $n_i^B$ be the number of ${{v}_{i}}$'s friends with property $A$, $B$, respectively. We then start to show how to integrate the ego network information for estimating the proportion of individuals with property $A$ in the population, $P_{A}^{*}$.
Assuming that the RDS process is conducted on a connected, undirected network with assumptions (i)-(iii) fulfilled, the probability that each node will be included in the sample, $\Pr ({v_i})$, will be proportional to its degree~\cite{ref17Salganik2004,ref18Heckathorn2007,ref19Volz2008}:

\begin{equation}
Pr ({{v}_{i}})\sim{\ }\frac{{{d}_{i}}}{\sum\nolimits_{j=1}^{N}{{{d}_{j}}}}, \label{eq3}
\end{equation}

where $N$ is the size of the population of interest.

Consequently, the probability that each link ${e_{i \to j}}$ will be selected to recruit a friend, $\Pr ({{e}_{i\to j}})$, depends on $\Pr ({v_i})$. Under the random recruitment assumption, we have:

\begin{equation}
Pr ({e_{i \to j}}) = \Pr ({v_i})\cdot\frac{1}{{{d_i}}} \sim \frac{1}{{\sum\nolimits_{j = 1}^N {{d_j}} }}, \label{eq4}
\end{equation}

that is, each link has the same probability of being selected via the RDS process. Consequently, the observed recruitment matrix $S$ is a random sample for the cross-group links of the network~\cite{ref17Salganik2004}.

The above are general inferences from a typical RDS process. Up to now, we can turn our attention to the egocentric data source. Let $\Pr (e_{i\to j}^{ego})$ be the probability that link ${e_{i \to j}}$ will be reported by ``ego'' ${v_i}$, since ${{e}_{i\to j}}$ is reported as long as ${{v}_{i}}$ is included in the sample, then:

\begin{equation}
\Pr (e_{i \to j}^{ego}) = \Pr ({v_i})\sim \frac{{{d_i}}}{{\sum\nolimits_{j = 1}^N {{d_j}} }}. \label{eq5}
\end{equation}

Consequently, to estimate the proportion of type ${e_{X \to Y}}{\text{ }}(X,Y \in \{ A,B\} )$ links in the population, $s_{XY}^{*}$, we can weigh the observed number of type $e_{X \to Y}^{ego}$ links by their inclusion probability to construct a \emph{generalized Hansen-Hurwitz estimator}~\cite{ref29Hansen1943}:

\begin{equation}
\hat{s}_{XY}^{ego}=\frac{\hat{N}_{XY}^{ego}}{\hat{N}_{XA}^{ego}+\hat{N}_{XB}^{ego}}=\frac{\sum\nolimits_{{{v}_{i}}\in X\bigcap U}^{{}}{\frac{n_{i}^{Y}}{{{d}_{i}}}}}{\sum\nolimits_{{{v}_{j}}\in X\bigcap U}^{{}}{\frac{n_{j}^{A}}{{{d}_{j}}}+\sum\nolimits_{{{v}_{j}}\in X\bigcap U}^{{}}{\frac{n_{j}^{B}}{{{d}_{j}}}}}}, \label{eq6}
\end{equation}

where $\hat N_{XY}^{ego} = \sum\nolimits_{{v_i} \in X \cap U}^{} {\frac{{n_i^Y}}{{{d_i}}}}$ is the weighted number of type ${e_{X \to Y}}$ links reported in the sample's ego networks.

Since the denominator in (6) can be rewritten as:

\begin{equation}
\sum\nolimits_{{{v}_{j}}\in X\bigcap U}^{{}}{\frac{n_{j}^{A}}{{{d}_{j}}}+\sum\nolimits_{{{v}_{j}}\in X\bigcap U}^{{}}{\frac{n_{j}^{B}}{{{d}_{j}}}}}=\sum\nolimits_{{{v}_{j}}\in X\bigcap U}^{{}}{(\frac{n_{j}^{A}+n_{j}^{B}}{{{d}_{j}}}})=\sum\nolimits_{{{v}_{j}}\in X\bigcap U}^{{}}{(\frac{{{d}_{j}}}{{{d}_{j}}}})={{n}_{X}}, \label{eq7}
\end{equation}

we have:

\begin{equation}
\hat s_{XY}^{ego} = \frac{1}{{{n_X}}} \cdot \sum\nolimits_{{v_i} \in X \cap U}^{} {\frac{{n_i^Y}}{{{d_i}}}}. \label{eq8}
\end{equation}

Note that in (8), the recruitment links are also counted as reported ego-alter links, and taking \autoref{fig1} as an example, ${e_{i \to j}}$ and ${e_{j \to i}}$ will be counted as $blue\to green$ type ego-alter link and $green \to blue$ type ego-alter link, separately.

Using $\hat{s}_{XY}^{ego}$ from (8) as an alternative to $S$, which is used in the $RDSI$ estimator, we can estimate $P_A^*$ by the same equation as (1). For the sake of clarity, the procedure for deriving (1) is replicated as follows:

In an undirected network, the number of cross-group links from $A$ to $B$ should equal the number of links from $B$ to $A$:

\begin{equation}
{N_A}\bar D_A^*s_{AB}^* = {N_B}\bar D_B^*s_{BA}^*. \label{eq9}
\end{equation}

where ${N_A} = N - {N_B}$ is the number of individuals of group $A$ in the population, and $\bar D_A^*$, $\bar D_B^*$ are average degrees for the two groups.

If we let $\hat{s}_{XY}^{ego}$ be the estimator of $s_{XY}^*$ and let ${\hat {\bar D}_X} = \frac{{{n_X}}}{{\sum\limits_{{v_i} \in X \cap U} {d_i^{ - 1}} }}$ be the estimator of ${\bar D_X^*}$ ($X,Y \in \{ A,B\} $), then $P_{A}^{*}$ can be estimated by:

\begin{equation}
{\hat P_A} = \frac{{\hat s_{BA}^{ego}{{\hat {\bar D}}_B}}}{{\hat s_{AB}^{ego}{{\hat {\bar D}}_A} + \hat s_{BA}^{ego}{{\hat {\bar D}}_B}}}{\text{ }{(RDSI^{ego})}}.\label{eq10}
\end{equation}

In all, the $RDSI^{ego}$ estimator uses the ego network data-based estimation of recruitment matrix, $\hat s_{XY}^{ego}$, instead of the observed $S$ used in $RDSI$. There are at least two advantages to using $\hat s_{XY}^{ego}$ rather than ${s_{XY}}$:

First, the sample size for inferring $\hat{s}_{XY}^{ego}$, is considerably larger than that for ${s_{XY}}$, reducing random error and making the estimates more reliable;

Second, in real RDS practice, respondents can hardly recruit their friends randomly~\cite{ref4Kogan2011, ref8Tomas2011, ref25deMello2008}, which leads to unknown bias and error for the representativeness of ${s_{XY}}$. $\hat s_{XY}^{ego}$, on the other hand, takes all of an ego's links into consideration, and consequently avoids this problem. Even the inclusion probability for a node may be shifted away from $\Pr ({v_i})$ when there are non-random recruitments; as we will see in section 4, $\hat s_{XY}^{ego}$ can greatly reduce estimate bias and error for such violation of assumption.

Note also the the implementation of $RDSI^{ego}$ does not necessarily require each respondent $i$ to list each of her/his alters' property: since degree is always collected in RDS, an estimated proportion of friends with a certain property $A$, $r_i^A$, would be enough to get to know the number of alters from group $A$, $r_i^Ad_i$.

\section{Simulation study design}

\subsection{Network data}

In this paper we use both an anonymized empirical social network and simulated networks to evaluate the performance of the newly proposed estimator. The empirical network, previously analyzed in~\cite{ref7Lu2012, ref23Lu2012, ref30Rybski2009}, comes from
the Nordic region's largest and most active web community for homosexual, bisexual, transgender, and queer persons. Nodes of the network are website members who identify themselves as homosexual males, and links are friendship relations defined as two nodes adding each other on their ``favorite list'', based on which they maintain their contacts and send messages. Only nodes and links within the giant connected component are used for this study, yielding a network of size $N = 16082$, and average degree ${\bar D^*} = 6.74$. Four dichotomous properties from users' profiles have been studied: \emph{age} (born before 1980), county (live in Stockholm, \emph{ct}), civil status (married, \emph{cs}), and profession (employed, \emph{pf}). The population value of group proportion ($P_A^*$), cross-group link probability ($s_{AB}^*$), homophily, and activity ratio, are listed in \autoref{table 1}.

\begin{table}[h]
\centering
\caption{\label{table 1}Basic statistics for variables in the MSM network}
\smallskip
{

\begin{tabular}{ccccc}
\hline
variable&$P_A^*{\text{ }}(\% )$&$s_{AB}^*$&Homophily&Activity ratio\\
\hline

$age$&77.8&0.13&0.40&1.05\\
$ct$&38.8&0.30&0.50&1.22\\
$cs$&40.4&0.57&0.05&0.97\\
$pf$&38.2&0.54&0.13&1.21\\
\hline
\end{tabular}
}
\end{table}

\emph{Homophily}, quantified as ${h_A} = 1 - s_{AB}^*/P_B^*$, is the probability that nodes connect with their friends who are similar to themselves rather than randomly. If the homophily of a property is 0, it means that all nodes are connected to their friends purely randomly, regardless of this property; if the homophily is 1, it means that all nodes with a particular property are connected to friends with the same property. \emph{Activity ratio}, is the ratio of mean degree for group $A$ to group $B$, $w=\bar{D}_{A}^{*}/\bar{D}_{B}^{*}$. Previous studies have found that homophily and activity ratio are two critical factors that may affect the performance of RDS estimators~\cite{ref6Gile2010}. Generally, the larger the homophily or difference between a group's mean degrees, the larger will be the bias and variance of the estimates. The various levels of homophily and activity ratio of the four variables in the MSM network provides a rich test base for RDS estimators. For example, the homophily for the county is 0.50, which means that members who live in Stockholm form links with members who also live in Stockholm 50\% of the time, while they form links randomly among all cities (including Stockholm) the remaining 50\% of the time. The civil status has a very low level of homophily, indicating that edges are formed as if randomly among other members, regardless of their marital status.

To systematically evaluate the effect of homophily and activity ratio on the performance of RDS estimators, we have also generated a set of simulated networks with ${h_A} \in [0,\ 0.5]$ and ${w} \in [0.5,\ 2.5]$ based on the KOSKK model, which is among the best social network models that can produce most realistic network structure with respect to degree distributions, assortativity, clustering spectra, geodesic path distributions, and community structure, and the like~\cite{Toivonen2009, Kumpula2007}. These networks are configured with population size $N=10000$, average degree $\bar D^*=10$, and population value $P_A^*=30\%$ (see Appendix for details). 


\subsection{Study design}

Based on the MSM network and artificial KOSSK networks, RDS processes are then simulated and the sample proportions and estimates are compared with population value to evaluate the accuracy of different estimators. In particular, we consider the following aspects:

\emph{Sample size}: we set the sample size to 500.

\emph{Sampling without replacement (SWOR)}: alike most empirical RDS studies, nodes are not allowed to be recruited again if they have already been in the sample.

\emph{Number of seeds and coupons}: following~\cite{ref6Gile2010}, we consider two scenarios: 6 seeds with 2 coupons, contributing to 500 respondents from 6 waves, and 10 seeds with 3 coupons, contributing to 500 respondents from 4 waves. However, we do not find significant difference for both settings in simulations and thus choose to show results with 6 seeds and 2 coupons.

\emph{Random and differential recruitment}: one of the assumptions that is most unlikely to be met in real life is that participants randomly recruit peers. For example, respondents may tend to recruit people who they think will benefit most from the RDS incentives~\cite{ref4Kogan2011}. In a study of MSM in Campinas City, Brazil~\cite{ref25deMello2008}, participants were reported most often to recruit close peers or peers they believed practiced risky behaviors. In~\cite{ref6Gile2010,ref7Lu2012,ref8Tomas2011}, it has been shown that all current RDS estimators would generate bias when the outcome variables are related to the tendency of such non-random distribution of coupons among respondents' personal networks (differential recruitment).

To test the robustness of the new estimator, we consider both scenarios. Let $p_A^{diff} \in [0,\ 1]$ be the probability that individuals from group $A$ are $p_A^{diff}$ times more likely to be recruited by both group $A$ members and group $B$ members, then $p_A^{diff}=0$ corresponds to random recruitment, when coupons are randomly distributed to respondents' friends, and $p_A^{diff}=1$ corresponds to the extreme case scenario that both group $A$ members and group $B$ members are twice as likely to recruit peers of type $A$, which would largely oversample both individuals from group $A$ and the proportion of recruitment links toward group $A$, ${s_{AA}}$ and ${{s}_{BA}}$.

\emph{Reporting error about degree and ego networks}: the new estimator requires respondents to report ego network information, bringing a new challenge in RDS. We simulate reporting error in two stages of a RDS process: first, when a respondent reports his or her degree, any alters of type $A$ or $B$ will be missed and not reported with probability $p_A^{miss}$ or  $p_B^{miss}$, respectively; second, when the composition of an ego network is reported, any alters of type $A$ will be misclassified as type $B$ with probability $p_{A \mapsto B}^{error}$, and any alters of type $B$ will be misclassified as type $A$ with probability $p_{B \mapsto A}^{error}$ vice versa.

\emph{RDS estimators}: since previous studies have suggested that sample composition may sometimes be an even better approximation of $P_A^*$ than traditional RDS estimators~\cite{ref9McCreesh2012,ref10Goel2010}, in addition to $RDSI$ and $RDSI^{ego}$, we also include the raw sample composition in the analysis. The $RDSII$ estimator in our simulations provides estimates with little difference to $RDSI$ and is thus not presented separately.

Since we are interested in generating feasible population estimates by information only collected within the RDS sample, the newly developed estimators that require known population parameters~\cite{ref22Gile2011,ref23Lu2012,ref24Krista2011} are thus beyond the purpose of this study and are excluded from comparison.

Four measurements are then carried out after the RDS simulations: the \emph{Bias}, which is the absolute difference between the average estimate and population value, $\left| {\frac{{\sum\nolimits_{i = 1}^m {es{t_i}} }}{m} - P_A^*} \right|$ or $\left| \frac{\sum\nolimits_{i=1}^{m}{es{{t}_{i}}}}{m}-s_{AB}^{*} \right|$, where $es{t_i}$ is the estimate from the ${i^{th}}$ simulation and $m$ the number of simulation times; the \emph{Standard Deviation} (SD) of estimates; the \emph{Root Mean Square Error} (RMSE), $\sqrt {\frac{{\sum\nolimits_{i = 1}^m {{{(es{t_i} - P_A^*)}^2}} }}{m}} $; and lastly, the \emph{Percentage} an estimator outperforms the rest in all simulations: ${P^{best}} = \frac{{times{\text{ }}the{\text{ }}estimator{\text{ }}gives{\text{ }}closest{\text{ }}estimate{\text{ }}to{\text{ }}s_{AB}^*{\text{ }}or{\text{ }}P_A^*}}{m}$.

All simulations were repeated 10,000 times, and seeds were excluded from the calculation of estimates in this study.

\section{Results}

\subsection{Random and differential recruitment}
\subsubsection{Estimates of network link types}

The difference between $RDSI$ and $RDSI^{ego}$ lies in the estimation of the recruitment matrix $S$. 
As a first step, we therefor simulate the RDS process with random recruitment ($p_A^{diff}=0$) and differential recruitment ($p_A^{diff}=1$) and then estimate the proportion of type ${e_{A \to B}}$ links in the population, $s_{AB}^{*}$, by both the raw sample recruitment proportion, ${s_{AB}}$, and the proposed ego-network-based estimator, $\hat s_{AB}^{ego}$, for all four variables in the MSM network, \emph{age}, \emph{ct}, \emph{cs} and \emph{pf}, respectively.

\begin{figure}[ht]
 \begin{center}
 \centerline{\includegraphics[width=0.7\textwidth]{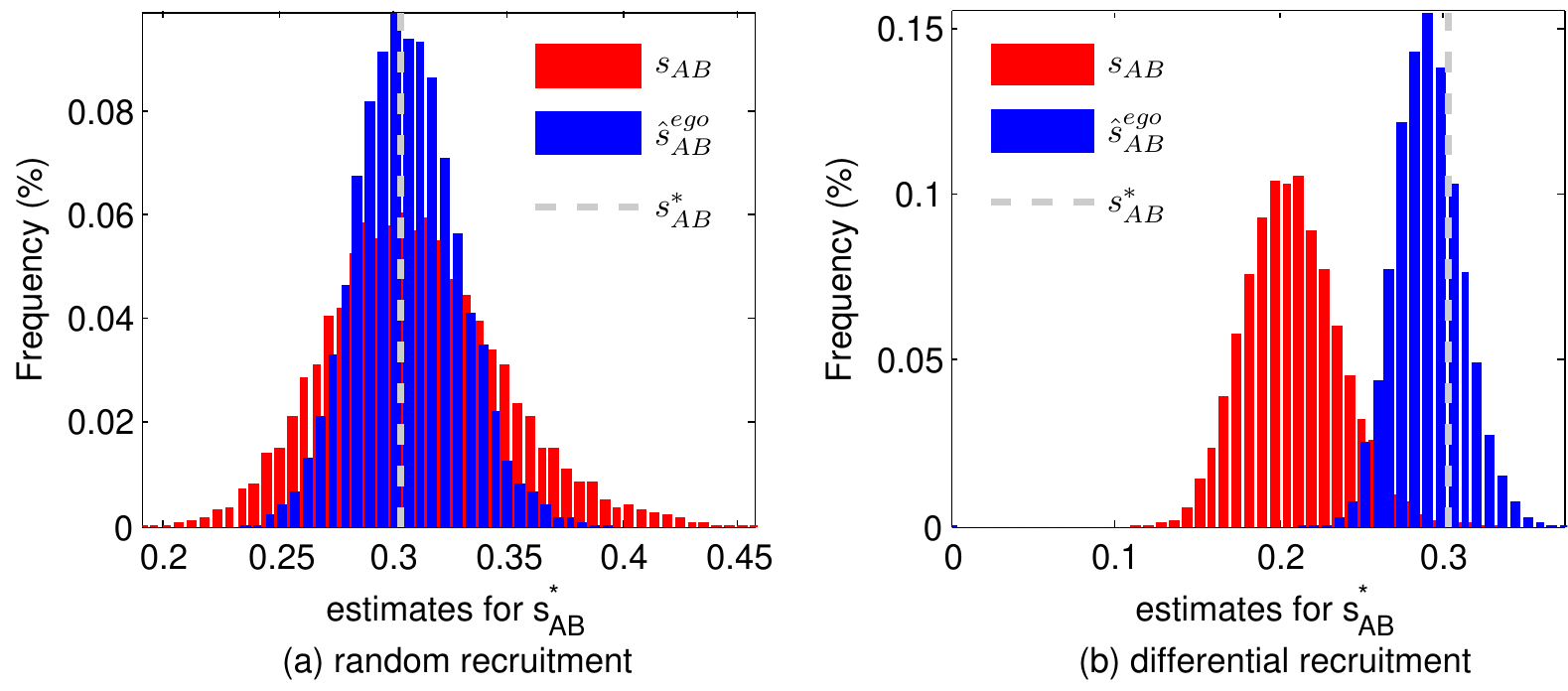}}
 \caption{Distribution of RDS estimates for $s_{AB}^*(ct)$. Dashed line shows population value $s_{AB}^{*}(ct)$. (a) Participants recruit respondents randomly among their friends, $p_A^{diff}=0$; (b) Participants are two times more likely to recruit friends of type $A$ than friends of type $B$,  $p_A^{diff}=1$.}\label{fig2}
 \end{center}
 \end{figure}

An example of the simulation results for \emph{ct} is presented in \autoref{fig2}. Clearly, when the random recruitment assumption is fulfilled (\autoref{fig2}(a)), both ${s_{AB}}$ and $\hat s_{AB}^{ego}$ are unbiased. 
Estimates by $\hat s_{AB}^{ego}$ peak more closely to $s_{AB}^{*}$ and have less variance than ${s_{AB}}$ (SD = 0.02 compared to 0.04, see \autoref{table 2}. The difference between ${s_{AB}}$ and $\hat s_{AB}^{ego}$ becomes more evident when RDS is implemented with differential recruitment. We can see from \autoref{fig2}(b) that when peers who live in Stockholm are two times more likely to be recruited by their friends, the raw sample recruit proportion is largely undersampled (Bias=0.09), while $\hat{s}_{AB}^{ego}$ still provides robust estimates (Bias=0.01) with less variance (SD=0.02). If we compare the performance of estimates for each simulation under random recruitment, $\hat s_{AB}^{ego}$ is 70\% times closer to $s_{AB}^*$ than ${s_{AB}}$. Under differential recruitment, almost all $\hat{s}_{AB}^{ego}$ estimates (${P^{best}} = 0.98$) outperform ${s_{AB}}$.

Simulation results for estimates of all variables are summarized in \autoref{table 2}. The conclusions are similar to those above. $\hat s_{AB}^{ego}$ gives less bias, SD, RMSE, and gives for most instances closer estimates, regardless of homophily and activity ratios. The precision of ${s_{AB}}$ depends largely on the random recruitment assumption; the bias and RMSE of ${s_{AB}}$ are a maximum of 0.01 and 0.04 for all simulation settings when peers are randomly recruited, while the maximum bias and RMSE all increase to 0.13 when differential recruitment happens. $\hat s_{AB}^{ego}$, on the other side, shows great robustness to violation of this assumption. The maximum bias and RMSE for all variables are less than 0.02 and 0.03, respectively.

Regarding ${{P}^{best}}$, $\hat s_{AB}^{ego}$ produces estimates that are closer to the true population value $s_{AB}^*$ 62\% to 74\% of the time when sampling is with random recruitment; when sampling with differential recruitment, ${P^{best}}$ increases to 77\%$\sim$100\%, revealing the superior performance of $\hat s_{AB}^{ego}$ over ${s_{AB}}$.
\begin{center}
\scalebox{0.85}{
\begin{threeparttable}[h]

\caption{\label{table 2}Statistics of estimates for $s^{*}_{AB}$ by ${s_{AB}}$ and $\hat s_{AB}^{ego}$}


\begin{tabular}{ccccccc}
\toprule
\multicolumn{2}{c}{}&\multicolumn{2}{c}{Bias (standard deviation)}&{}&\multicolumn{2}{c}{RMSE ($P^{best}$)}\\
\cline{3-4}
\cline{6-7}

\multicolumn{2}{l}{\emph{Random recruitment}}&{$s_{AB}$}&{$\hat s_{AB}^{ego}$}&{}&{$s_{AB}$}&{$\hat s_{AB}^{ego}$}\\
\cline{3-4}
\cline{6-7}

\multirow{4}{*}{\tabincell{c}{seed=6\\coupon=2\\SWOR}}
&\emph{age}&.00 (.03)&.00\tnote{*} (.03\tnote{*} )&{}&.03 (.37)&.03\tnote{*} (.63\tnote{*} )\\
&\emph{ct}&.01 (.04)&.00\tnote{*} (.02\tnote{*} )&{}&.04 (.30)&.02\tnote{*} (.70\tnote{*} )\\
&\emph{cs}&.00 (.04)&.00\tnote{*} (.02\tnote{*} )&{}&.04 (.26)&.02\tnote{*} (.74\tnote{*} )\\
&\emph{pf}&.00 (.04)&.00\tnote{*} (.02\tnote{*} )&{}&.04 (.26)&.02\tnote{*} (.74\tnote{*} )\\


\midrule

\multicolumn{7}{l}{\emph{Differential recruitment}}\\			

\multirow{4}{*}{\tabincell{c}{seed=6\\coupon=2\\SWOR}}
&\emph{age}&.04 (.03)&.01\tnote{*} (.03\tnote{*} )&{}&.05 (.16)&.03\tnote{*} (.84\tnote{*} )\\
&\emph{ct}&.09 (.03)&.01\tnote{*} (.02\tnote{*} )&{}&.10 (.02)&.02\tnote{*} (.98\tnote{*} )\\
&\emph{cs}&.13 (.04)&.02\tnote{*} (.02\tnote{*} )&{}&.13 (.00)&.03\tnote{*} (1.0\tnote{*} )\\
&\emph{pf}&.13 (.03)&.02\tnote{*} (.02\tnote{*} )&{}&.13 (.00)&.02\tnote{*} (1.0\tnote{*} )\\


\bottomrule

%

\end{tabular}

\begin{tablenotes}
\item [*] {corresponding statistic is better than the other estimator.}
\end{tablenotes}

\end{threeparttable}

}
\end{center}

\subsubsection{Estimates of population compositions}
The superiority of $\hat s_{AB}^{ego}$ over ${s_{AB}}$ shown in the above section suggests that the $RDSI^{ego}$ estimator should also give less bias and error than $RDSI$. To confirm this, we compare the simulation results of $RDSI$, and $RDSI^{ego}$ to estimate population proportions on both the MSM network and the KOSKK networks.

\begin{figure}[ht]
 \begin{center}
 \centerline{\includegraphics[width=0.7\textwidth]{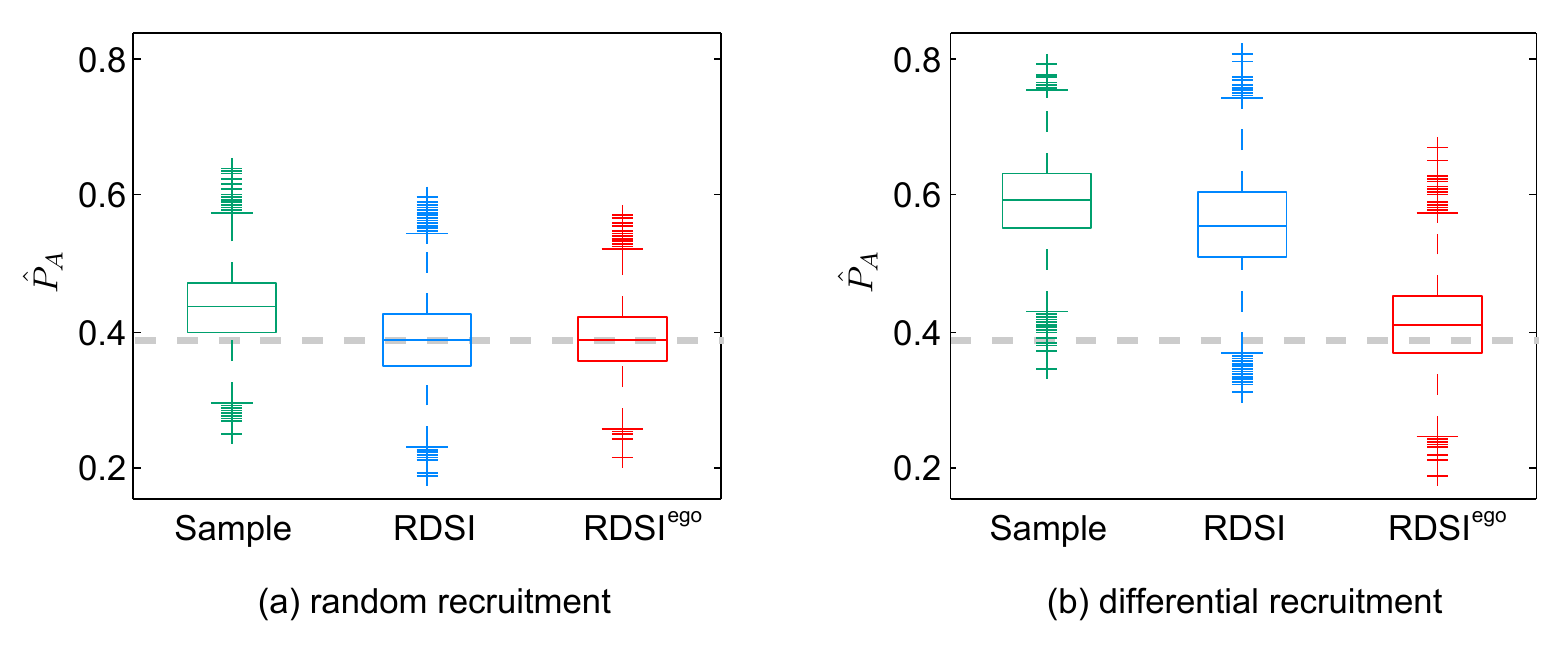}}
 \caption{RDS estimates for $P_{ct}^*$. Dashed line is of population value $P_{ct}^*$. (a) Participants recruit respondents randomly among their friends, $p_A^{diff}=0$; (b) Participants are two times more likely to recruit friends of type $A$ rather than friends of type $B$, $p_A^{diff}=1$.}\label{fig3}
 \end{center}
 \end{figure}

First, we take the estimates of $P_A^*$ for \emph{ct} as an example. The result is presented as boxplots in \autoref{fig3}, where the median (middle line), the 25th and 75th percentiles (box) and outliers (whiskers) are shown. When  $p_A^{diff}=0$, there is on average an oversample of individuals who live in Stockholm (0.05) in the raw sample; however, if adjusted, $RDSI$ and $RDSI^{ego}$ all give unbiased estimates (\autoref{fig3}(a)). 
When  $p_A^{diff}=1$, i.e., respondents are twice as likely to recruit friends from Stockholm rather than friends from other counties, the improvement in estimates by $RDSI^{ego}$ becomes much more significant. While the sample composition/$RDSI$ has a bias of 0.20/0.17 and RMSE of 0.21/0.18, the bias for $RDSI^{ego}$ is only 0.02 and RMSE is 0.06 (\autoref{fig3}(b)). Another notable finding is that the number of times an estimator provides the closest estimate is almost equal between sample composition and $RDSI$ under random recruitment (${P^{best}} = 0.28$ for sample composition and ${P^{best}} = 0.29$ for $RDSI$, see \autoref{table 3}), implying that even when the RDS sample is collected under ideal conditions, the traditional adjusted population estimates may perform as poorly as the raw sample proportion. $RDSI^{ego}$, by contrast, produces estimates closest to $P_A^*$\ 43\% of the time. For sampling with differential recruitment, $RDSI^{ego}$ is far superior to the other estimators, with ${P^{best}} = 0.93$.

The above conclusions are similar for all other variables (see \autoref{table 3}): when $p_A^{diff}=0$, both $RDSI$ and $RDSI^{ego}$ have little bias, while $RDSI^{ego}$ generates less SD and RMSE, and provides the closest estimates than the rest estimators 10\% more often. It is interesting to compare ${P^{best}}$ of sample composition with the rest of the estimators; $RDSI^{ego}$ always has larger ${P^{best}}$ for all variables except \emph{cs}, which has low homophily and a close to one activity ratio. $RDSI$, by contrast, cannot consistently outperform the sample composition. It has almost the same probability of providing the closest estimate to $P_{A}^{*}$ as the sample composition for \emph{ct}, and is even less likely to be better when estimating \emph{age} and \emph{cs}. $RDSI^{ego}$ again becomes dominant when the sampling is done with differential recruitment. The bias ranges in [0.00, 0.02] and RMSE in [0.04, 0.07], while for sample composition and $RDSI$ the bias and RMSE are much larger, [0.07, 0.20] and [0.09, 0.21], respectively.
\begin{center}\scalebox{0.85}{
\begin{threeparttable}[h]
\caption{\label{table 3}Statistics of estimates for $P_A^*$ by sample mean, $RDSI$ and $RDSI^{ego}$}

\begin{tabular}{ccccccccc}
\toprule
\multicolumn{2}{c}{}&\multicolumn{3}{c}{Bias (standard deviation)}&{}&\multicolumn{3}{c}{RMSE ($P^{best}$)}\\
\cline{3-5}
\cline{7-9}

\multicolumn{2}{l}{\emph{Random recruitment}}&sample&$RDSI$&$RDS{I^{ego}}$&{}&sample&$RDSI$&$RDS{I^{ego}}$\\
\cline{3-5}
\cline{7-9}

\multirow{4}{*}{\tabincell{c}{seed=6\\coupon=2\\SWOR}}
&\emph{age}&.01 (.06)&.00\tnote{*} (.07)&.00 (.06\tnote{*} )&{}&.06 (.37)&.07 (.23)&.06\tnote{*} (.40\tnote{*} )\\
&\emph{ct}&.05 (.05)&.00 (.06)&.00\tnote{*} (.05\tnote{*} )&{}&.07 (.28)&.06 (.29)&.05\tnote{*} (.43\tnote{*} )\\
&\emph{cs}&.01 (.03\tnote{*} )&.00 (.04)&.00\tnote{*} (.03)&{}&.03\tnote{*} (.51\tnote{*} )&.04 (.17)&.03 (.32)\\
&\emph{pf}&.05 (.03\tnote{*} )&.00\tnote{*} (.04)&.00 (.03)&{}&.05 (.20)&.04 (.32)&.03\tnote{*} (.48\tnote{*} )\\


\midrule

\multicolumn{9}{l}{\emph{Differential recruitment}}\\			

\multirow{4}{*}{\tabincell{c}{seed=6\\coupon=2\\SWOR}}
&\emph{age}&.09 (.05\tnote{*} )&.08 (.06)&.02\tnote{*} (.07)&{}&.10 (.10)&.10 (.12)&.07\tnote{*} (.79\tnote{*} )\\
&\emph{ct}&.20 (.06)&.17(.07)&.02\tnote{*} (.06\tnote{*} )&{}&.21 (.00)&.18 (.06)&.06\tnote{*} (.93\tnote{*} )\\
&\emph{cs}&.12 (.03\tnote{*} )&.13 (.05)&.02\tnote{*} (.04)&{}&.13 (.00)&.14 (.04)&.04\tnote{*} (.96\tnote{*} )\\
&\emph{pf}&.18 (.03\tnote{*} )&.13 (.05)&.02\tnote{*} (.04)&{}&.18 (.00)&.14 (.05)&.04\tnote{*} (.95\tnote{*} )\\
%

\bottomrule

%

\end{tabular}

\begin{tablenotes}
\item [*] {corresponding statistic is better than other estimators.}
\end{tablenotes}

\end{threeparttable}

}
\end{center}

\begin{figure}[h!]
 \begin{center}
 \centerline{\includegraphics[width=0.65\textwidth]{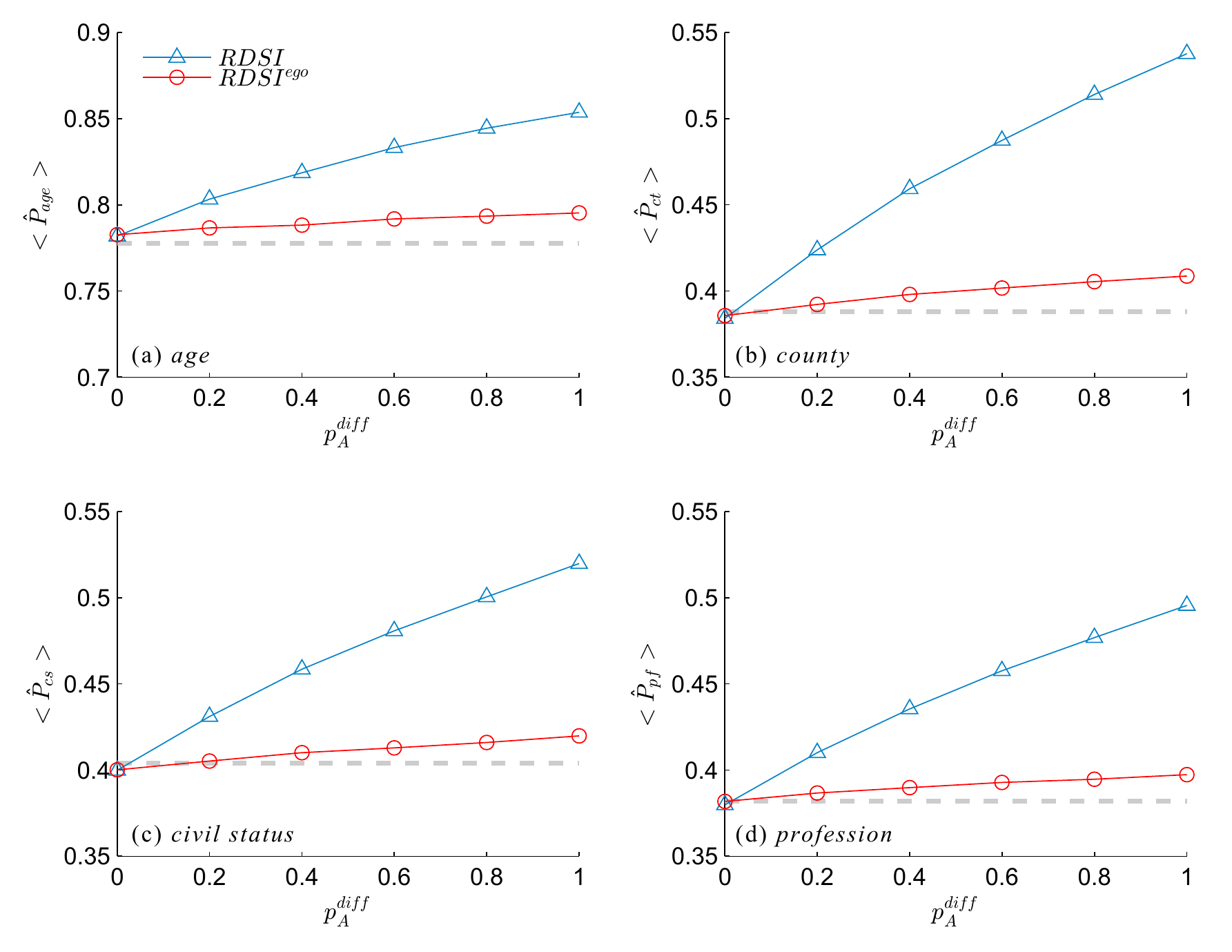}}
 \caption{Average estimates of $RDSI$ and $RDSI^{ego}$ on the MSM network with varying level of differential recruitment.}\label{fig4}
 \end{center}
 \end{figure}

To better understand the robustness of $RDSI^{ego}$ to differential recruitment, we simulate RDS processes on the MSM network with $p_A^{diff}$ varying from 0 to 1. The average estimates for the four variables are shown in \autoref{fig4}. 
While the bias of $RDSI$ increases progressively with $p_A^{diff}$, $RDSI^{ego}$ shows a clear resistance over different levels of differential recruitment. Additionally, we can see that the magnitude of bias of $RDSI$ does not depend solely on either the homophily or activity ratio, implying that, without the collection of ego network information, more sophisticated modifications are needed for $RDSI$ to adapt differential recruitment.

\begin{figure}[h!]
 \begin{center}
 \centerline{\includegraphics[width=0.7\textwidth]{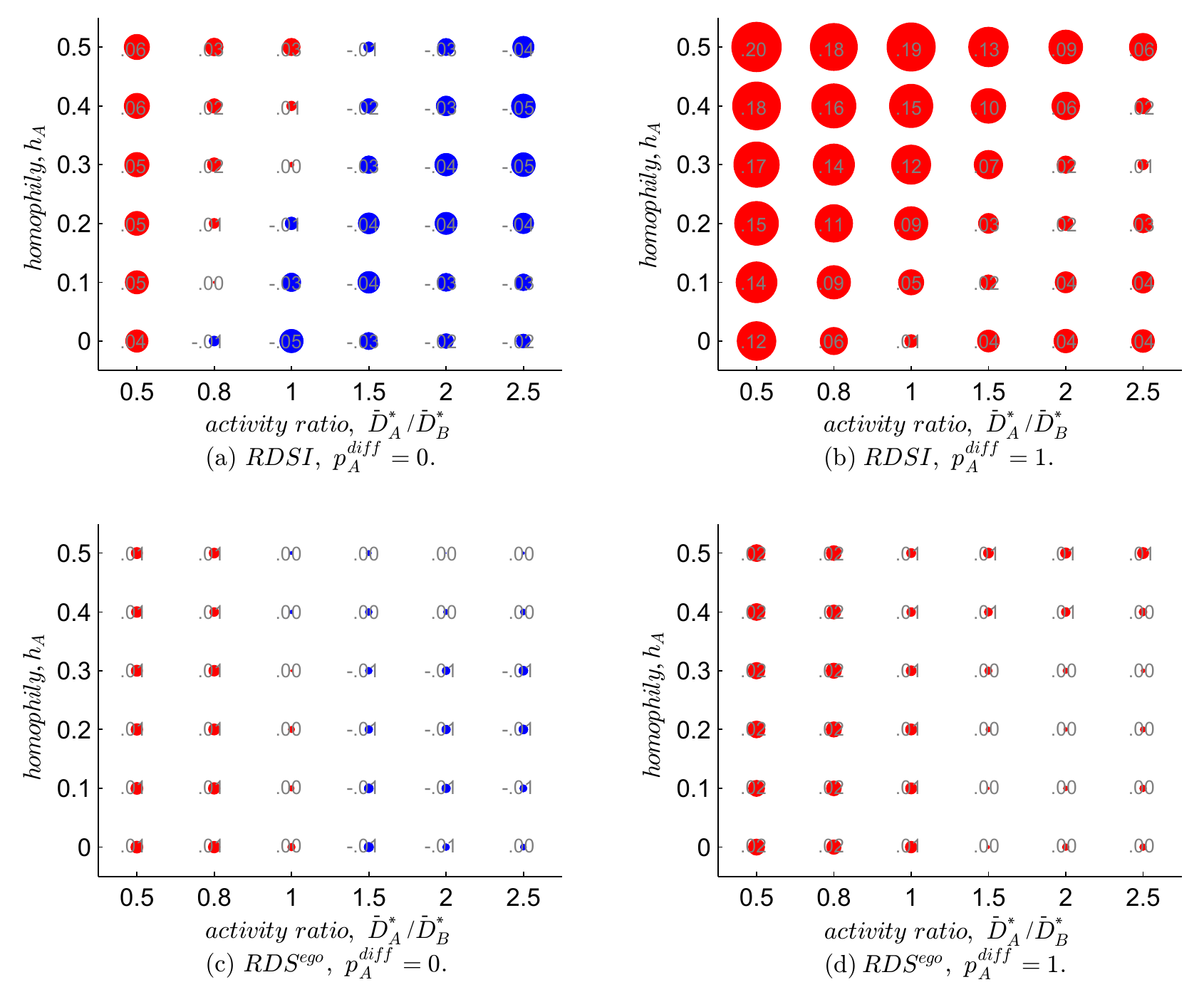}}
 \caption{Bias of $RDSI$ and $RDSI^{ego}$ on KOSKK networks with random recruitment (a, c) and differential recruitment (b, d) .}\label{fig5}
 \end{center}
 \end{figure}

The complexity of joint effect of homophily and activity ratio is more evident for RDS estimates on the KOSKK networks, as shown in \autoref{fig5}, where the biases of both $RDSI$ and $RDSI^{ego}$ are shown for networks with different levels of homophily ($h_A\in [0,\ 0.5]$) and activity ratio ${w} \in [0.5,\ 2.5]$.

Generally, bias increases with homophily and difference between average degrees, however, these effects are mixed with impact of other network structural properties, for example the community structure resulted by the KOSKK model, making networks with certain combinations of $h_A$ and ${w}$ least biased. $RDSI^{ego}$ shows resistance over all these structural effects: when $p_A^{diff}=0$, the bias for $RDSI$ ranges from 0.00 to 0.06, while for $RDSI^{ego}$, this range is only $[0.00, 0.01]$; when $p_A^{diff}=1$, the maximum bias for $RDSI$ goes up to 0.20, while the maximum bias for $RDSI^{ego}$ stays around 0.02.
\subsection{Degree reporting error}

With the superior performance observed from the above section, we will from this section focus on $RDSI^{ego}$ and evaluate factors that may bring extra sources of biases.

The degree reporting error parameters $p_A^{miss}$ and $p_B^{miss}$, capture the fact that in social network surveys, especially surveys targeting hidden populations, individuals in the target population may not be identified by their friends and would thus be miscounted when a respondent reports the personal network size~\cite{Salganik2011, ref23Lu2012}. This reporting error will not only affect the estimates of average degree, but further bias the estimate of the recruitment matrix in $RDSI^{ego}$, $\hat s_{XY}^{ego} (X,\ Y \in {A, B})$.

\begin{figure}[h!]
 \begin{center}
 \centerline{\includegraphics[width=0.65\textwidth]{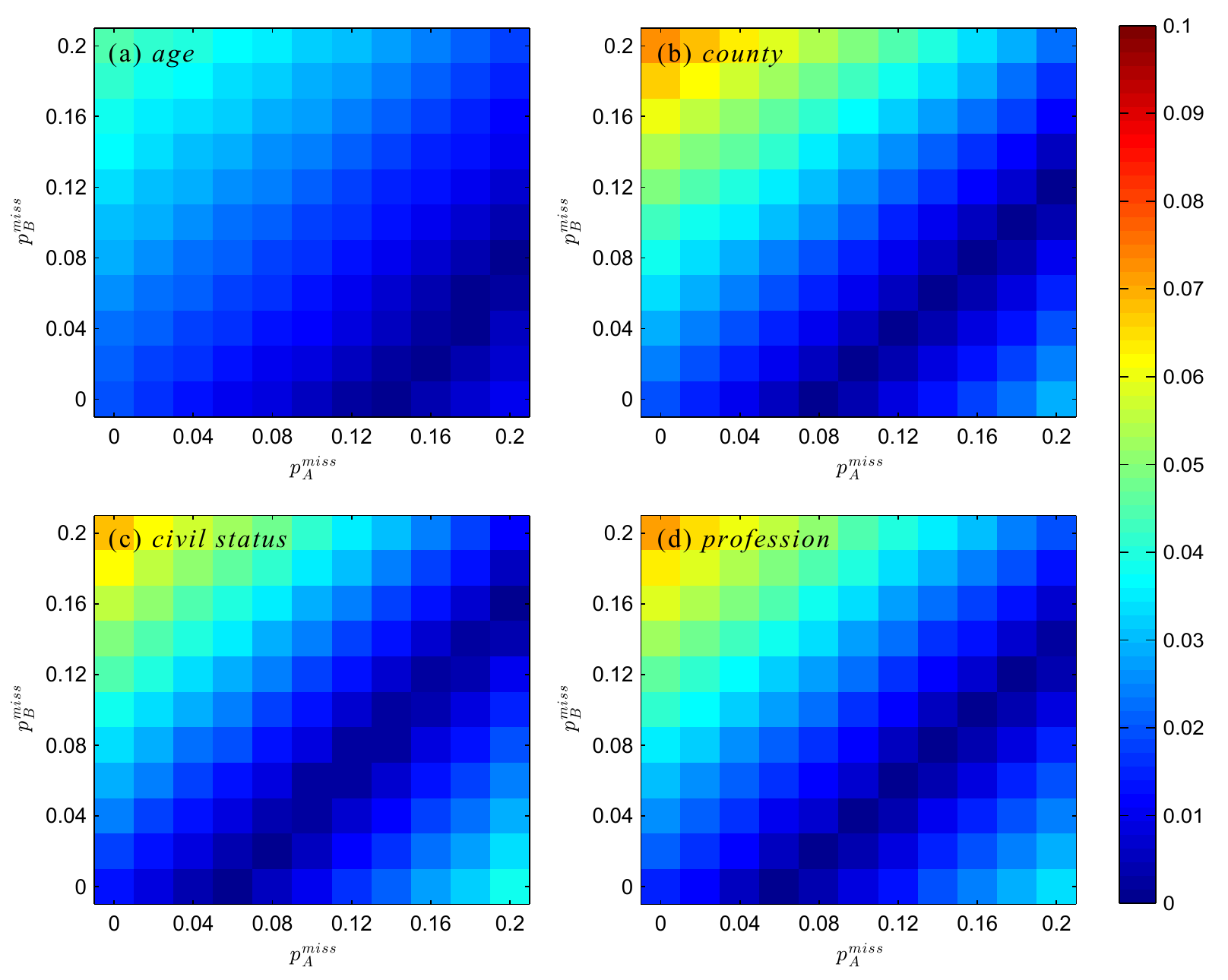}}
 \caption{Bias of $RDSI^{ego}$ on the MSM network with differential recruitment and \emph{degree reporting error}.}\label{fig6}
 \end{center}
 \end{figure}

\begin{figure}[t]
 \begin{center}
 \centerline{\includegraphics[width=0.65\textwidth]{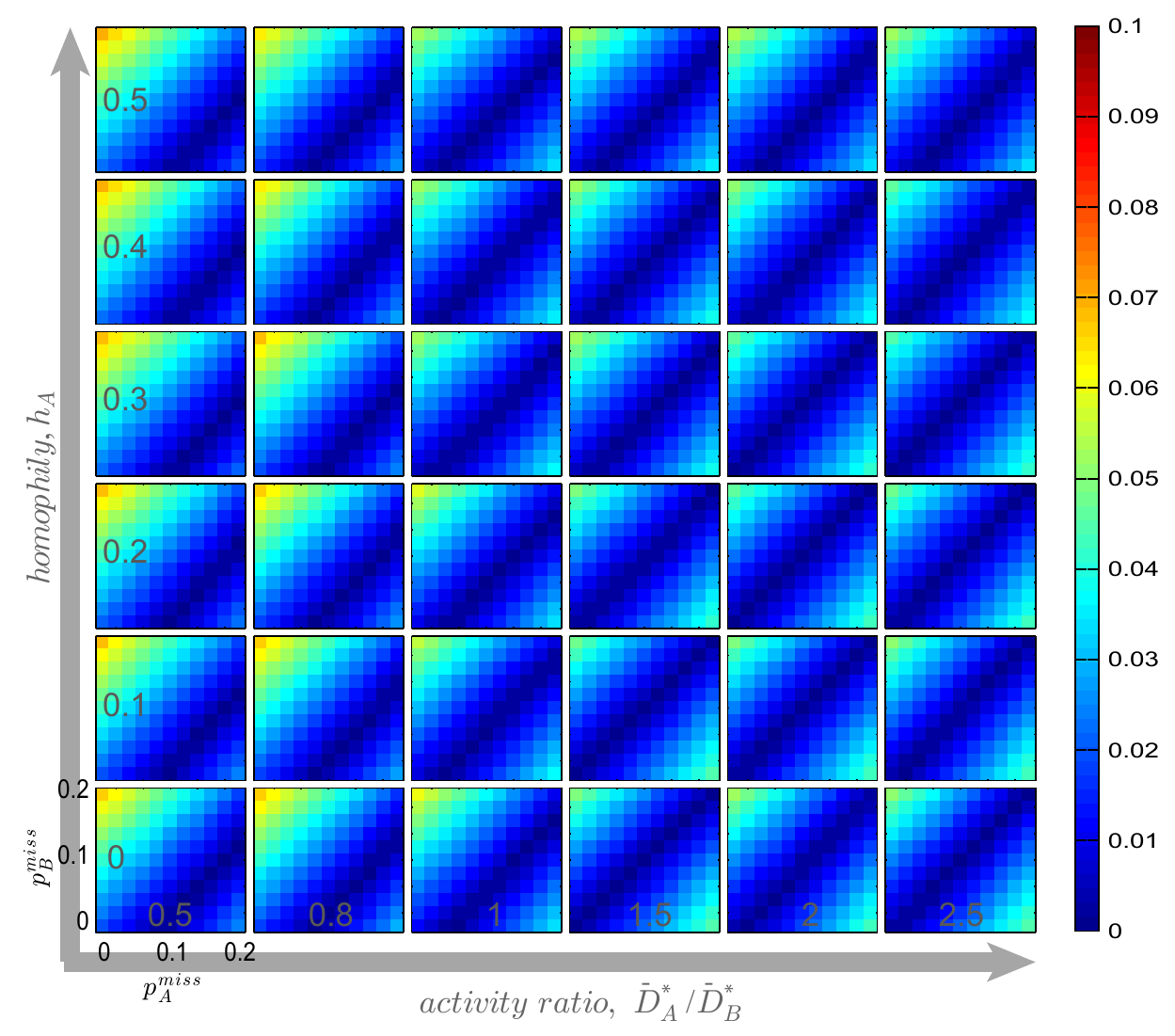}}
 \caption{Bias of $RDSI^{ego}$ on KOSKK network with differential recruitment and \emph{degree reporting error}.}\label{fig7}
 \end{center}
 \end{figure}

We simulate RDS with degree reporting error $p_A^{miss}\in [0,\ 0.2]$ and $p_B^{miss}\in [0,\ 0.2]$, that is, a maximum of 20\% friends with property $A$ or $B$ may be unidentified as the target population. To account for the absolute worst case scenario, differential recruitment ($p_A^{diff}=1$) is also included in the simulation. Results are presented in \autoref{fig6} for the MSM network and \autoref{fig7} for KOSKK networks.

Surprisingly, on both the MSM network and KOSKK networks, even with 20\% of all alters being miscounted, the biases of $RDSI^{ego}$ range mostly within $[0.00, 0.05]$ with a few exceptions. The worst case scenario occurs when 20\% of all alters from one group are missed in the reported degree, while none from the other group is missed, with the maximum bias around 0.07. When miscounted alters are less than 10\%, most configurations of $[p_A^{miss},\ p_B^{miss}]$ produce biases less than 0.04.

We can also see a symmetric effect of $p_A^{miss}$ and $p_B^{miss}$, the bias maintains on the same level as long as the two parameters change in the same direction. This effect was previously examined in \cite{ref7Lu2012}, where the degree reporting error was modeled as unawareness of existing relationships. These findings implies that the magnitude of bias resulted by degree reporting error is much less than the error itself, since the increase of reporting error on one group can ``compensate'' reporting error on the other group; tolerable bias would be expected when the reporting error is limited.

It is worth noting that the biases analyzed here are outcomes of RDS simulations with ``extreme'' differential recruitment. We have also ran simulations with random recruitment ($p_A^{diff}=0$), which generate similar patterns (e.g., the symmetric effect, where the maximum bias occurs) with smaller biases, see Appendix \autoref{Sfig6} and \autoref{Sfig7}.

\subsection{Ego network reporting error}
\begin{figure}[ht]
 \begin{center}
 \centerline{\includegraphics[width=0.65\textwidth]{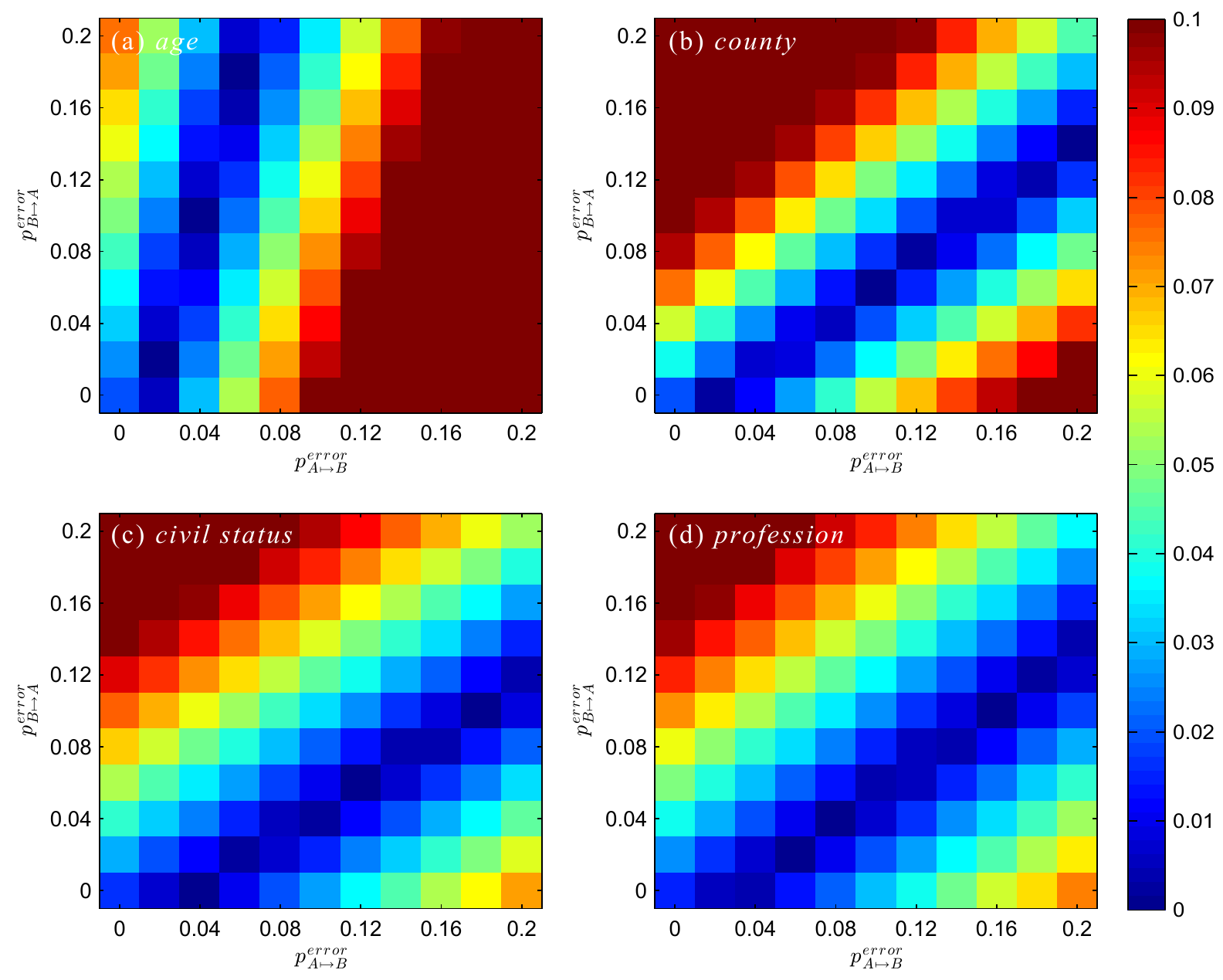}}
 \caption{Bias of $RDSI^{ego}$ on the MSM network with differential recruitment and \emph{ego network reporting error}.}\label{fig8}
 \end{center}
 \end{figure}

\begin{figure}[ht]
 \begin{center}
 \centerline{\includegraphics[width=0.65\textwidth]{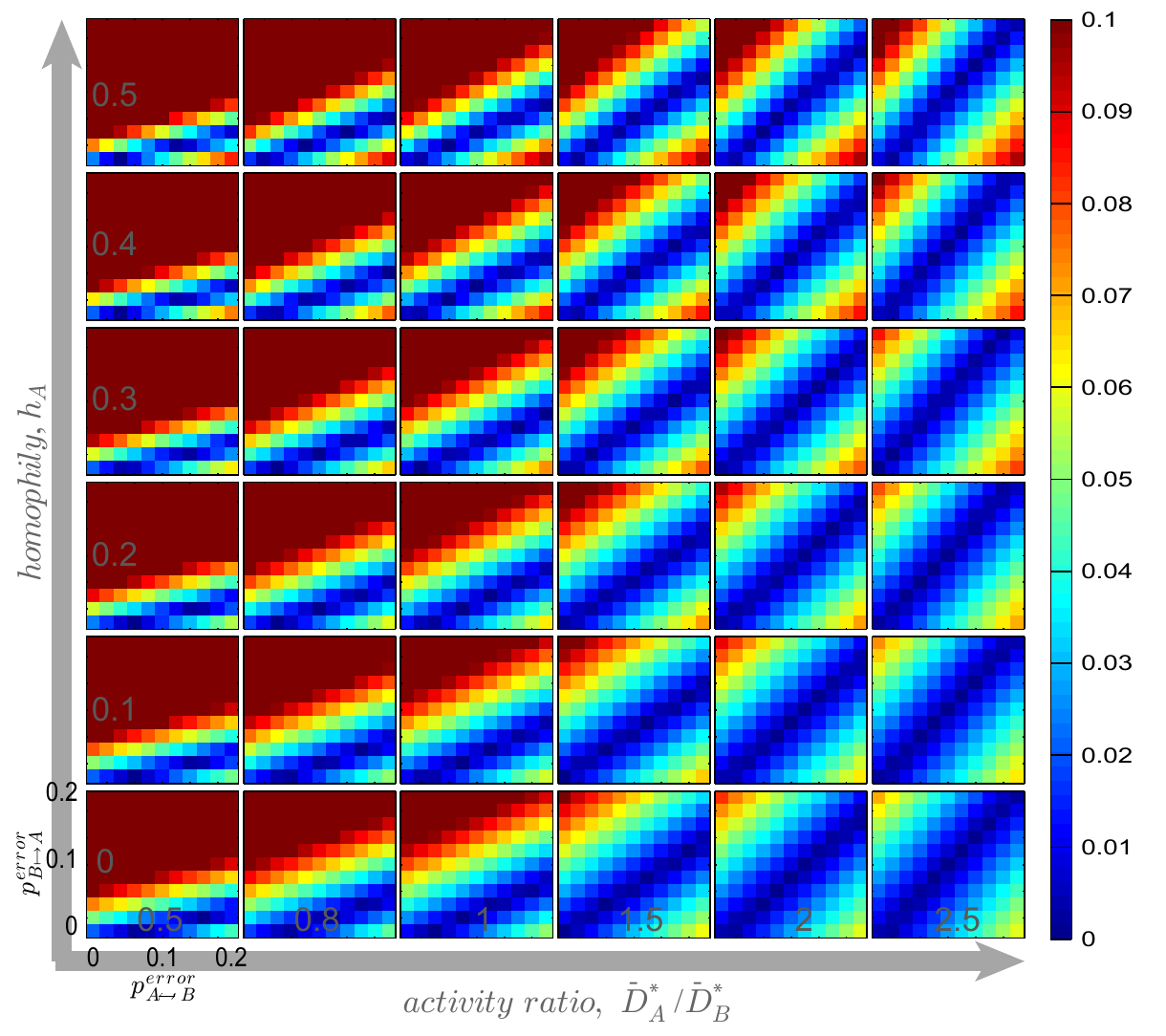}}
 \caption{Bias of $RDSI^{ego}$ on KOSKK network with differential recruitment and \emph{ego network reporting error}.}\label{fig9}
 \end{center}
 \end{figure}

Another reporting error related to the implementation of $RDSI^{ego}$, is that even when individuals fulfilling the sample inclusion criteria are correctly identified, their characteristics, especially for sensitive variables such as HIV status and sexual preference, may be incorrectly reported by their friends. By varying $p_{A \mapsto B}^{error}$ and $p_{B \mapsto A}^{error}$ from 0, when the composition of ego networks are accurately reported, to 0.2, when 20\% of the alters are misclassified, we run simulations on the MSM networks and KOSKK networks, to evaluate the sensitivity of $RDSI^{ego}$ to the reporting error in ego network compositions. Similar to the previous section, we use differential recruitment and set $p_A^{diff}=1$. Results are shown in \autoref{fig8} and \autoref{fig9}.

Contrary to the robustness to degree reporting error, the $RDSI^{ego}$ estimator is much more sensitive to the ego network reporting error on both the MSM network and KOSKK networks. On the MSM network, the bias readily excesses 0.1 as long as $p_A^{diff}>0.1$ and $p_B^{diff}<0.1$ for \emph{age}, and $p_A^{diff}<0.1$ and $p_B^{diff}>0.1$ for \emph{ct}. The biases for the other two variables with less homophily are relatively smaller, as long as the misclassification error for alters of both groups is less than 10\%.

Given $p_{A \mapsto B}^{error} \in [0, 0.2]$ and $p_{B \mapsto A}^{error}\in [0, 0.2]$, the ego network reporting error on KOSKK networks produces much larger bias for networks with low activity ratios ($w \leq 1$). And the increase of $p_{B \mapsto A}^{error}$ is apparently more harmful than the increase of $p_{A \mapsto B}^{error}$. This effect is due to the fact that when $w\leq 1$, a large amount of alters for respondents in the RDS sample are from group $B$ (note also $P_B^*=0.7$), a small probability of misclassifying $B$ alters as $A$ alters will result in a large absolute number of over-reported $A$ alters in the end, making $RDSI^{ego}$ generate estimates much higher than the true population value $P_A^*$. For this reason, variables with high activity ratios, on the other hand, are less sensitive to the network reporting error.

The above reasoning can also be verified with estimates for \emph{age} on the MSM network, which has a relatively balanced activity ratio ($w=1.05$), but a population proportion of 70\%. Therefore, reporting error regarding the group with higher population proportion and activity ratio will result in substantial amount of misclassified alters in the ego networks and greatly affect the estimates.

Simulations with random recruitment have also been carried out, however the ego network reporting error seems to be the dominant factor driving estimate error for $RDS^{ego}$, no significant reduce of bias is observed, see Appendix \autoref{Sfig8} and  \autoref{Sfig9}.

\section{Conclusion and discussion}
Ego network data has been collected for decades and exists largely in sociological surveys~\cite{ref28Tom2012,ref33Handcock2010,ref34Newman2003,ref35Mizruchi2006,ref36Marsen2002,ref37Hanneman}; the RDS sampling mechanism further makes it possible to collect ``\emph{linked-ego network}'' data. By combining RDS recruitment trees with ego networks, this study developed a new estimator, $RDSI^{ego}$, for RDS studies. Given that participants can accurately report the composition of their personal networks, this estimator has superior performance over traditional RDS estimators. Most importantly, $RDSI^{ego}$ shows strong robustness to differential recruitment, a violation of the RDS assumptions that may cause large bias and estimation error and is not under the control of the researchers. Evaluation studies on our simulated KOSKK networks also show that $RDSI^{ego}$ performs consistently well on networks with varying homophily, activity ratio, and community structures. 

The limitation of $RDSI^{ego}$ is rooted in the need to collect ego network data. Many RDS studies are designed for use among hidden populations, who may be reluctant to share certain private information with their friends. Consequently, the proposed method is primarily suited for less sensitive variables, which the respondent can be expected to know about his contacts. Such information may for example include socio-demographic variables (e.g., gender, age groups, profession, marital status, etc.) for which survey methods on how to design and collect ego network data has been extensively studied~\cite{ref39Kogovsek2005,ref40Matzat2010,ref41Burt1984,ref42Martin2004}. Additionally, certain variables, e.g. drug use, may be highly sensitive in the general population but may not be at all be so in an IDU population.

By modeling the difficulty in understanding of personal network composition as degree reporting error and ego network reporting error, which quantify the level of mutual knowledge about studied variables shared with friends, we have showed that even with 20\% of alters being unidentified, $RDSI^{ego}$ was still able to produce estimates with bias less than 0.05 most of the time. On the other hand, $RDSI^{ego}$ is sensitive to the error of misclassifying alters. If 20\% of alters from one group is mistakenly reported as belonging to the other group, estimate bias can exceed 0.1 when the probability of misclassifying members of one group is substantially larger than misclassification of members in the other group (e.g., $p_{A \mapsto B}^{error} \gg p_{B \mapsto A}^{error}$). Fortunately, the result shows that when the studied variables only related to a small proportion of alters, that is, if $P_A^*$ is low and $w$ is relatively small, the increase of error in misclassifying $A$ as $B$ members will have a small influence on the bias. Consequently, for many sensitive variables surveyed in RDS studies, if the reporting error of a low prevalence trait (e.g., HIV status) is mainly ``false negatives'', e.g., alters with HIV are reported as healthy friends since they are reluctant to reveal this information to their egos, estimates with small bias are still expected to be able to achieve.

There are other interesting findings from this study. First, the performance of $RDSI$, which has been used in most RDS studies so far, fails to outperform the sample composition in many simulation settings. Second, we propose in this paper a new bootstrap method for constructing confidence intervals (CIs) with $RDSI^{ego}$ (see Appendix). Simulations in this paper and recent studies~\cite{ref9McCreesh2012,ref10Goel2010,ref38Salganik2012}, has shown that the traditional bootstrapping method underestimates variance. However, the proposed bootstrap method in this paper is able to generate CIs that much better approximate the expected coverage rates and performs fairly consistent to variations of homophily, activity ratio and community structures of networks.

In summary, we have shown that, by combining the traditional RDS sampling design with collection of ego network data, population estimates can improve drastically. What's most important, since RDS is a chain-referral designed sampling strategy, once the sample is started from seeds, the distribution of coupons is largely out of the control of researchers, and non-random recruitment often occurs, which has been proved to generate large estimate bias and error~\cite{ref6Gile2010,ref7Lu2012,ref8Tomas2011,linusAIDS}. The robustness of $RDSI^{ego}$ to differential recruitment offers researchers the ability to largely reduce estimate error. Additionally, by comparing $\hat{S}^{ego}$ with the observed raw sample recruitment matrix $S$, the severity of differential recruitment may be assessed. For future RDS studies, we encourage ego network questions to be integrated with traditional RDS questionnaires along with the improved bootstrap procedure. Due to the limitations inherent in the collection of sensitive variables from stigmatized group, the new method may be better suited to less sensitive variables. This new method is also applicable to sampling problems in other fields~\cite{ref20Gjoka,ref43Lee2006,ref44Yoon2007}, such as sampling of internet contents from which the ego network data is more reliable and may be more efficiently retrieved.

\section*{Acknowledgements}
The author would like to thank Professor Fredrik Liljeros and Dr. Linus Bengtsson for helpful discussions. This work has been funded by China Scholarship Council (Grant No. 2008611091) and Riksbankens Jubileumsfond (The Bank of Sweden Tercentenary Foundation).

\section*{Appendix}
\subsection*{Appendix A: Generation process for KOSKK networks}

As one of the dynamical network evolution models, the KOSKK model utilizes network link weights to generate networks with key common feathers of social networks~\cite{Kumpula2007}: (i) skewed degree distribution, (ii) assortative mixing, (iii) high average clustering coefficient, (iv) small average shortest path lengths, and (v) community structures. In a comprehensive comparative study~\cite{Toivonen2009}, the KOSKK model was found to be one of the best social network models that can generate similar-to-real social network structures, among nodal attribute models, network evolution models as well as ERGM models.

In a KOSKK model, the network is initiated with $N$ nodes and zero edges, and then evolved with three mechanisms:

(i) \emph{Local attachment}. Select a node $i$ randomly, and choose one of $i$'s neighbor $j$ with probability ${w_{ij}}/\sum\nolimits_j {{w_{ij}}}$, where $w_{ij}$ is the weight on link $e_{ij}$. If $j$ has another neighbor apart from $i$, choose one of them (node $k$) with probability ${w_{jk}}/\sum\nolimits_k {({w_{jk}} - {w_{ij}})}$. If there is no link between $i$ and $k$, connect $k$ to $i$ with probability $p_{\Delta}$ and set $w_{ki}=w_0$. Increase link weight $w_{ij}$, $w_{jk}$, and $w_{ki}$ (if was already present) by $\delta$.

(ii) \emph{Global attachment}. Connect $i$ to a random node $l$ with probability $p_r$ (or with probability 1 if $i$ has no connections) and set $w_{il}=w_0$.

(iii) \emph{Node deletion}. Select a random node and with probability $p_d$ remove all of its connections.

\begin{figure}[h]
\centering
\includegraphics[scale=0.72]{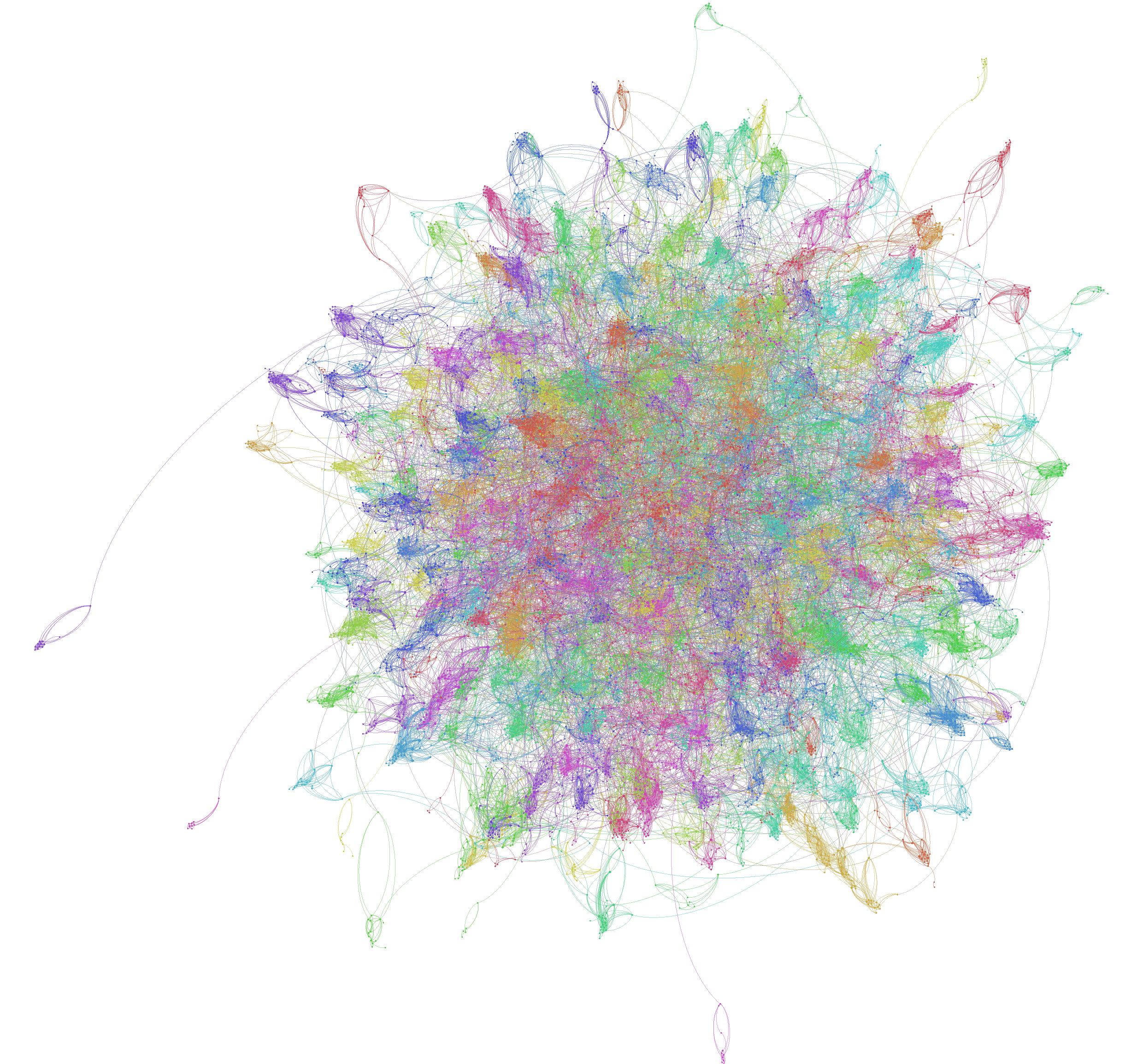}
\caption{Visualization of the KOSKK network generated with $\delta=0.6$, $\bar D^*=10$. }
\label{figKOSKK}
\end{figure}

With larger $\delta$, clearer community structures will be generated, as new links are created preferably through strong links. When $p_d$ is fixed, the average degree is obtained by adjusting $p_{\Delta}$ for each $\delta$. In our simulation, we set $N=10000$, $w_0=1$, $p_r=0.0005$, $p_d=0.001$, $\delta=0.6$, and the network average degree $\bar D^*=10$. The process was ran $10^8$ time steps to achieve stationary network characteristics. At the end of the process, a few nodes will be isolated due to the \emph{node deletion} step, we simply randomly link these nodes to the giant connected component to make sure all nodes in the network are connected. As $\delta$ is relatively large, the obtained network shows a clear community structure, see~\autoref{figKOSKK}.

Based on the above network, we then start the configuration of homophily and activity ratio. Let $w$ be the activity ratio of the current network and $w^*$ be the activity ratio we want to obtain. At the beginning, 30\% of the nodes are randomly selected and assigned with property $A$, the rest of nodes are then assigned with property $B$. If $w>w^*$, we randomly pick a node with property $A$, $i$, and a node with property $B$, $j$, if $d_i>d_j$, we then exchange the properties of the two nodes, i.e., $i$ becomes a $B$ node, and $j$ becomes a $A$ node. If $w<w^*$, we exchange the properties of $i$, $j$ only when $d_i<d_j$. The above process is repeated until $w=w^*$.

For each of the network configured with $w^*$, we use a rewiring process to adjust the homophily. Recall that the homophily is depended on the number of cross group links as ${h_A} = 1 - s_{AB}^*/P_B^*$, smaller $s_{AB}$ indicates high homophily. Let $h_A$ be the homophily of the current network and $h_A^*$ be the desired value, if $h_A>h_A^*$, we randomly pick two within group links $i\leftrightarrow j$, $k\leftrightarrow l$, with $i$, $j$ belonging to group $A$, and $k$, $l$ belonging to group $B$, and rewire them to $i\leftrightarrow k$, $j\leftrightarrow l$, to increase cross group links. Similarly, if $h_A<h_A^*$, we randomly pick two cross group links and rewire them to form two within group links. The above process is repeated until $h_A=h_A^*$.


\clearpage
\subsection*{Appendix B: Confidence interval estimation}
The precision of a sample estimate is usually enhanced by providing a confidence interval (CI), which gives a range within which the true population is expected to be found with some level of certainty. Due to the complex sample design of RDS, simple random sampling based CIs are generally narrower than expected~\cite{ref10Goel2010,ref16Heckathorn2002,ref21Salganik2006}. Consequently, bootstrap methods are used to construct CIs around RDS estimates.

The current widely used bootstrap procedure for RDS ($BS\textrm{-}origin$) was proposed by Salganik~\cite{ref21Salganik2006,ref32Volz2007}. In this procedure, respondents are divided into two groups depending on the property of their recruiters, that is, those who are recruited by $A$ nodes ($A_{rec}$), and those who are recruited by $B$ nodes ($B_{rec}$). Then the bootstrap starts by a randomly chosen respondent. If the respondent has property $A$, then the next respondent is randomly picked from $A_{rec}$, otherwise from $B_{rec}$. Such a procedure is repeated with replacement until the original RDS sample size is reached, then the RDS estimate is calculated based on the replicated sample. When $R$-replicated samples are bootstrapped, the resulting middle 90\%/95\% estimates from the ordered $R$ estimates are then used as the estimated CI.

We extend the $BS\textrm{-}origin$ in two different ways:

(a)	$BS\textrm{-}ego1$: we implement the same resampling procedure as with $BS\textrm{-}origin$; however, when each replicated sample is collected, $RDSI^{ego}$ is used to calculate the RDS estimate, rather than $RDSI$;

(b)	$BS\textrm{-}ego2$: we divide the sample into two groups depending on the property of the respondents, that is, those with property $A$ ($A_{set}$) and those with property $B$ ($B_{set}$). Then the bootstrap procedure is started with a randomly picked respondent. If the respondent has property $A$, then the probability of selecting the next respondent from $A_{set}$ or $B_{set}$, is $1 - \hat s_{AB}^{ego}$ and $\hat s_{AB}^{ego}$, respectively. If the respondent has property $B$, then the probability of selecting the next respondent from $A_{set}$ or $B_{set}$, is $\hat{s}_{BA}^{ego}$ and $1 - \hat s_{BA}^{ego}$, respectively. The above process is repeated until the same size as original sample is reached. $RDSI^{ego}$ is then used to calculate the RDS estimate for each replicated sample.

We expect that the modification in the bootstrap procedure of $BS\textrm{-}ego2$ by introducing the ego network data based estimate $\hat s_{AB}^{ego}$ and $\hat s_{BA}^{ego}$ can improve the performance of estimated CIs when the RDS is done with differential recruitment.

\begin{figure}[h!]
 \begin{center}
 \centerline{\includegraphics[width=0.7\textwidth]{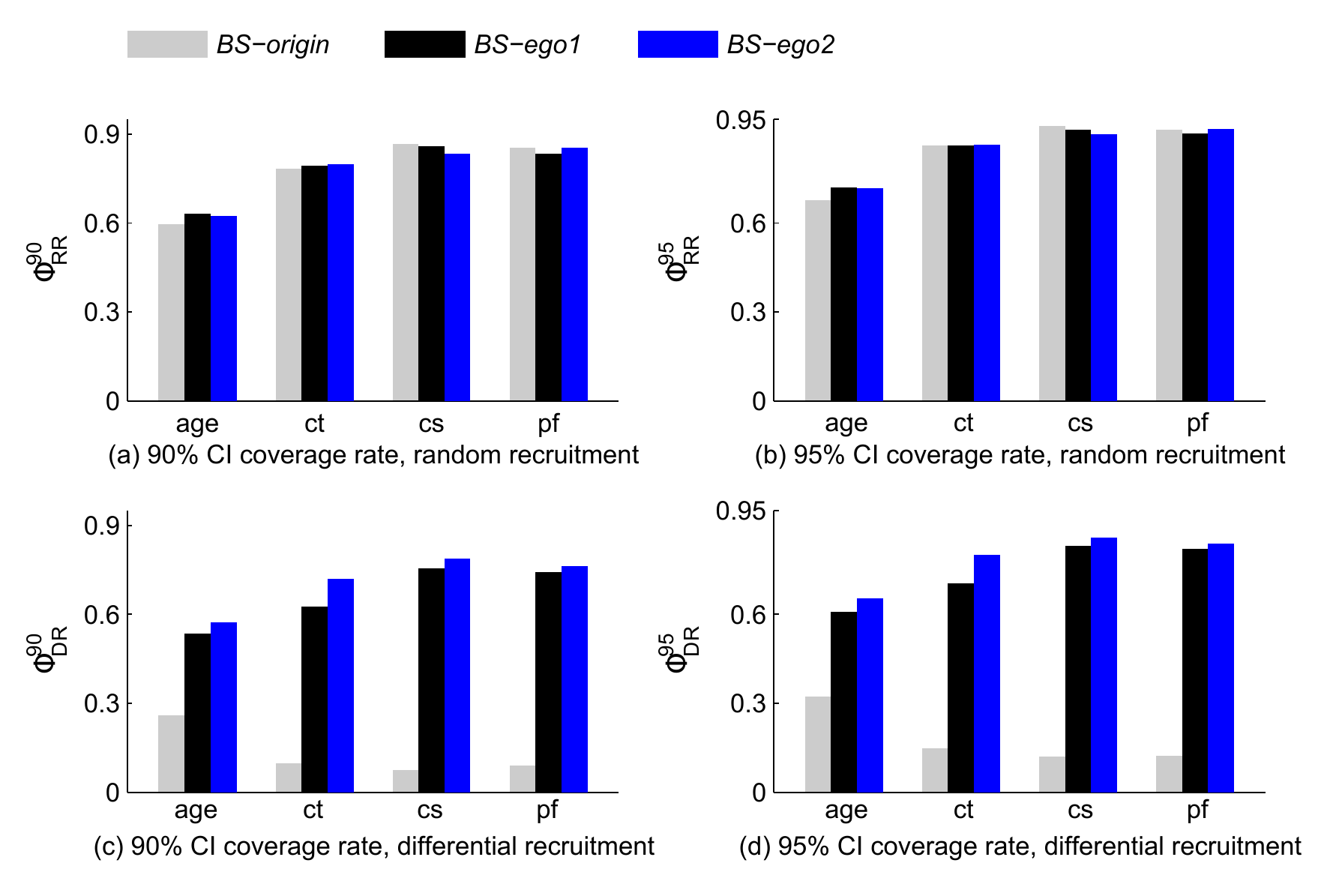}}
 \caption{Coverage rate of 90\% and 95\% confidence interval, by bootstrap procedure $BS\textrm{-}origin$, $BS\textrm{-}ego1$, and $BS\textrm{-}ego2$ on the MSM network.}\label{Sfig4}
 \end{center}
 \end{figure}

\begin{figure}[h!]
 \begin{center}
 \centerline{\includegraphics[width=0.7\textwidth]{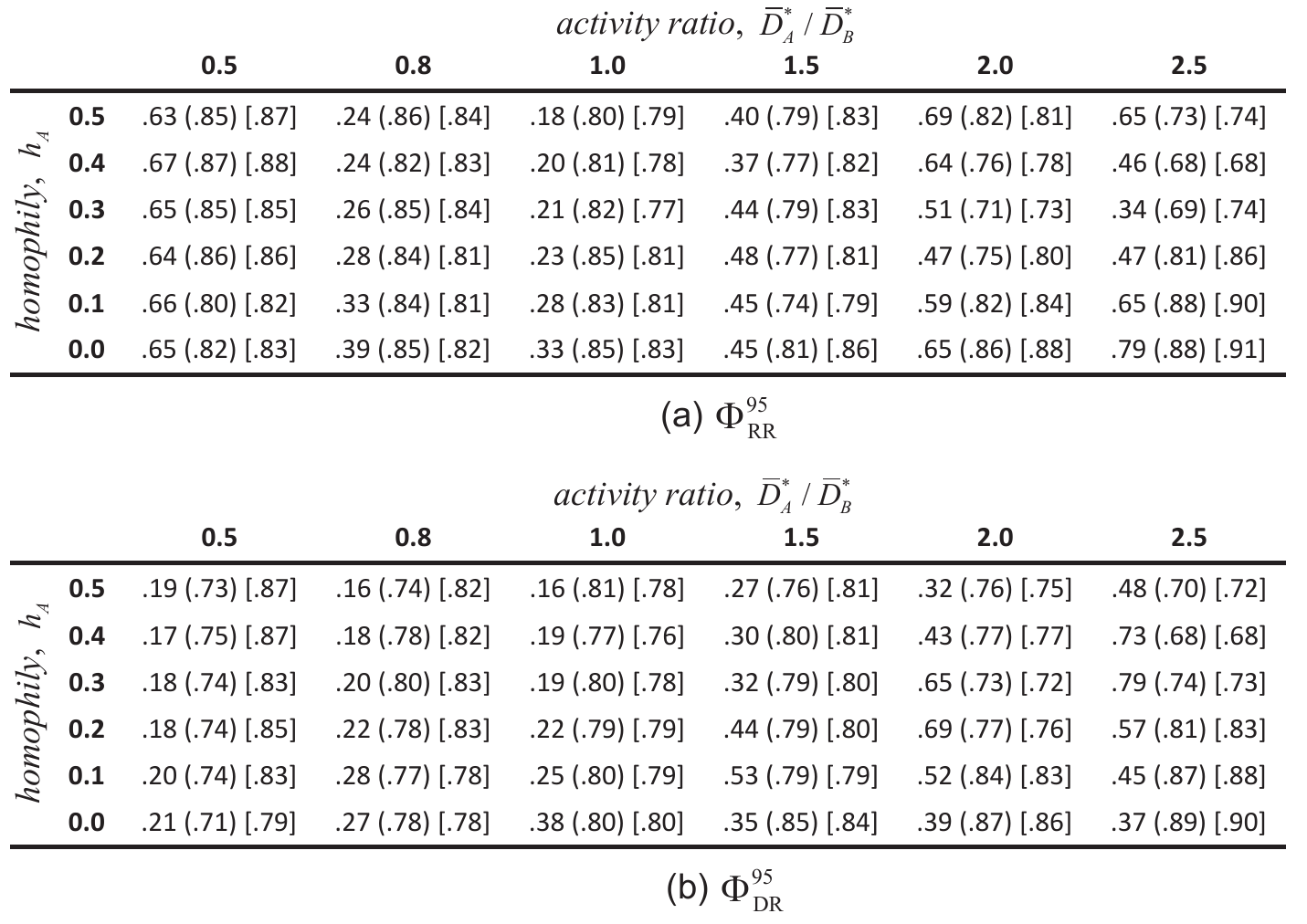}}
 \caption{Coverage rate of 95\% confidence interval, by bootstrap procedure $BS\textrm{-}origin$, ($BS\textrm{-}ego1$), and [$BS\textrm{-}ego2$] on KOSKK networks. (a) Random recruitment, $p_A^{diff}=0$; (b) Differential recruitment, $p_A^{diff}=1$.}\label{SfigCIKOSKK}
 \end{center}
 \end{figure}

Following~\cite{ref21Salganik2006}, we use simulations on both the MSM network and KOSKK networks to compare the performance of $BS\textrm{-}origin$, $BS\textrm{-}ego1$, and $BS\textrm{-}ego2$. For each variable, 1000 RDS samples are collected, and for each of these 1000 samples we construct the 90\% and 95\% CIs based on 1000 replicate samples drawn by the above bootstrap procedures. The proportion of times the generated confidence interval contains the true population value $P_A^*$ when sampling with random recruitment and differential recruitment (denoted as $\Phi _{RR}^{90}$, $\Phi _{DR}^{90}$, and $\Phi _{RR}^{95}$, $\Phi _{DR}^{95}$) is compared with different bootstrap methods and are presented in \autoref{Sfig4} and \autoref{SfigCIKOSKK} .

On the MSM network, when sampling with random recruitment, we can see from \autoref{Sfig4}(a), (b) that all three methods produce similar coverage rates for the tested variables. The coverage rate for \emph{age} is significantly smaller than the desired value for both $\Phi _{RR}^{90}$ and $\Phi _{RR}^{95}$, indicating that even under ideal conditions, the bootstrap-based CIs in RDS may be much narrower than expected. When the RDS is done with differential recruitment (\autoref{Sfig4}(c), (d)), the coverage rate of $BS\textrm{-}origin$ becomes extremely small and practically useless. This is because the $RDSI$ estimates are largely biased from the true population value when differential recruitment exists. The coverage rates of $BS\textrm{-}ego1$ and $BS\textrm{-}ego2$, on the other hand, are well above 50\% for all the four variables and therefore outperform $BS\textrm{-}origin$ in an absolute sense. In general, there is 5\%$\sim $10\% more coverage in $\Phi _{DR}^{90}$ and $\Phi _{DR}^{95}$ for $BS\textrm{-}ego2$ compared to $BS\textrm{-}ego1$, implying that the modified bootstrap procedure is more resistant to the violation of the random recruitment assumption in RDS. 

$BS\textrm{-}origin$ performs poorly on KOSKK networks for both sampling with random recruitment and sampling with differential recruitment, with a majority of 95\% coverage rates under 50\%. The $RDSI^{ego}$-based bootstrap methods, all produce coverage rates 20\%$\sim $60\% higher than $BS\textrm{-}origin$. When $p_A^{diff}=0$, there is no significant difference between $BS\textrm{-}ego1$ and $BS\textrm{-}ego2$, however, when $p_A^{diff}=1$, $BS\textrm{-}ego2$ is able to produce 8\%$\sim $14\% higher coverage rates than $BS\textrm{-}ego1$ in extreme cases ($w=0.5$).

It is worth noting that, even $BS\textrm{-}ego2$ shows superior performance over $BS\textrm{-}origin$ and is robustness to variations in network structure properties evaluated in this study (e.g., homophily, activity ratio, and the like.), the bootstrapped CIs rarely approach required coverage rates. On KOSKK networks, it is common that the 95\% coverage rates are 5\%$\sim $20\% lower than expected. Even the community structure in these networks may impede the performance of RDS estimates as well as the bootstrap methods, future work is needed to develop CI estimate methods with improved precision.
\clearpage

\subsection*{Appendix C: Supporting figures}

\begin{figure}[h!]
 \begin{center}
 \centerline{\includegraphics[width=0.605\textwidth]{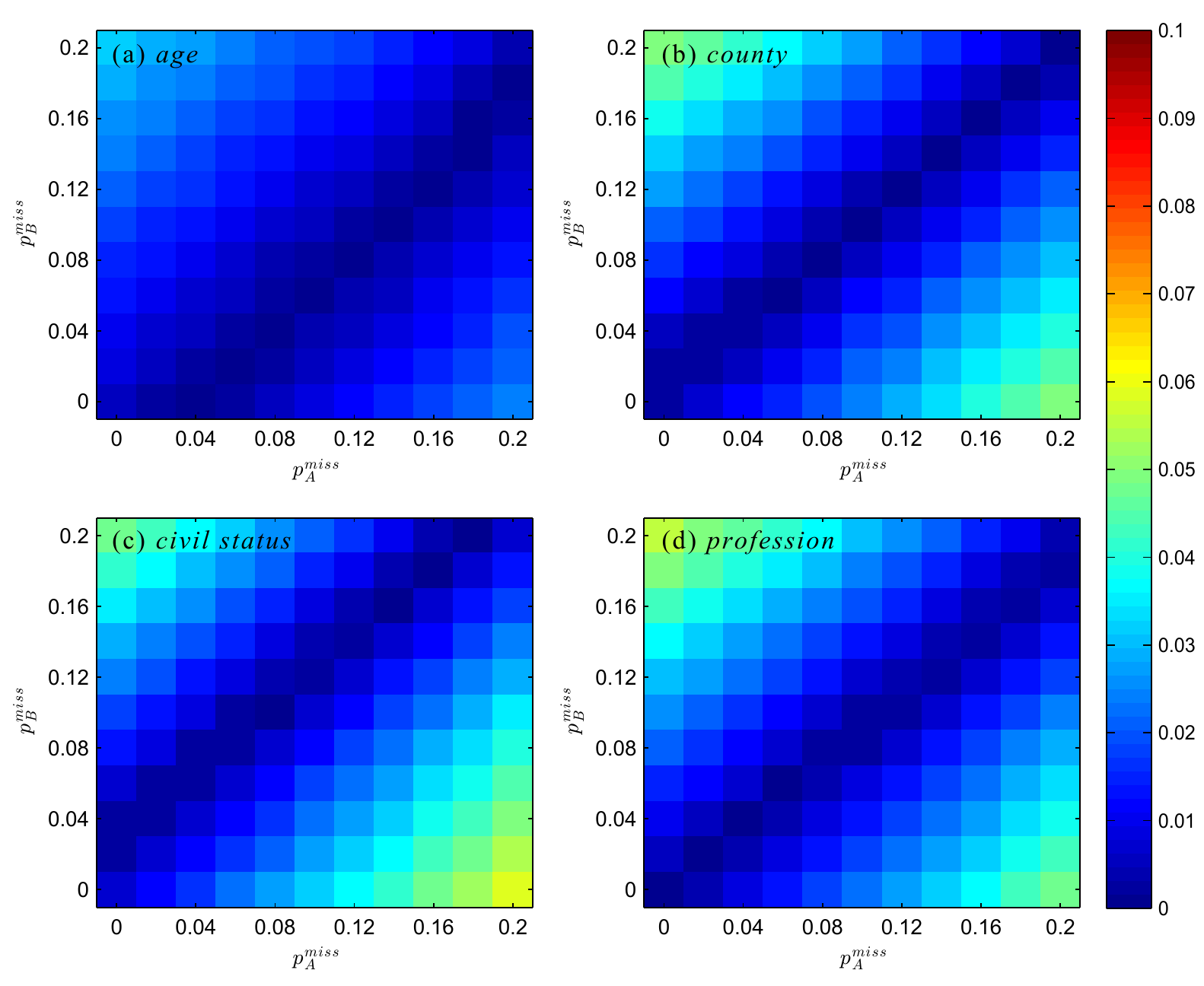}}
 \caption{Bias of $RDSI^{ego}$ on the MSM network with random recruitment and \emph{degree reporting error}.}\label{Sfig6}
 \end{center}
 \end{figure}

  \begin{figure}[h!]
 \begin{center}
 \centerline{\includegraphics[width=0.60\textwidth]{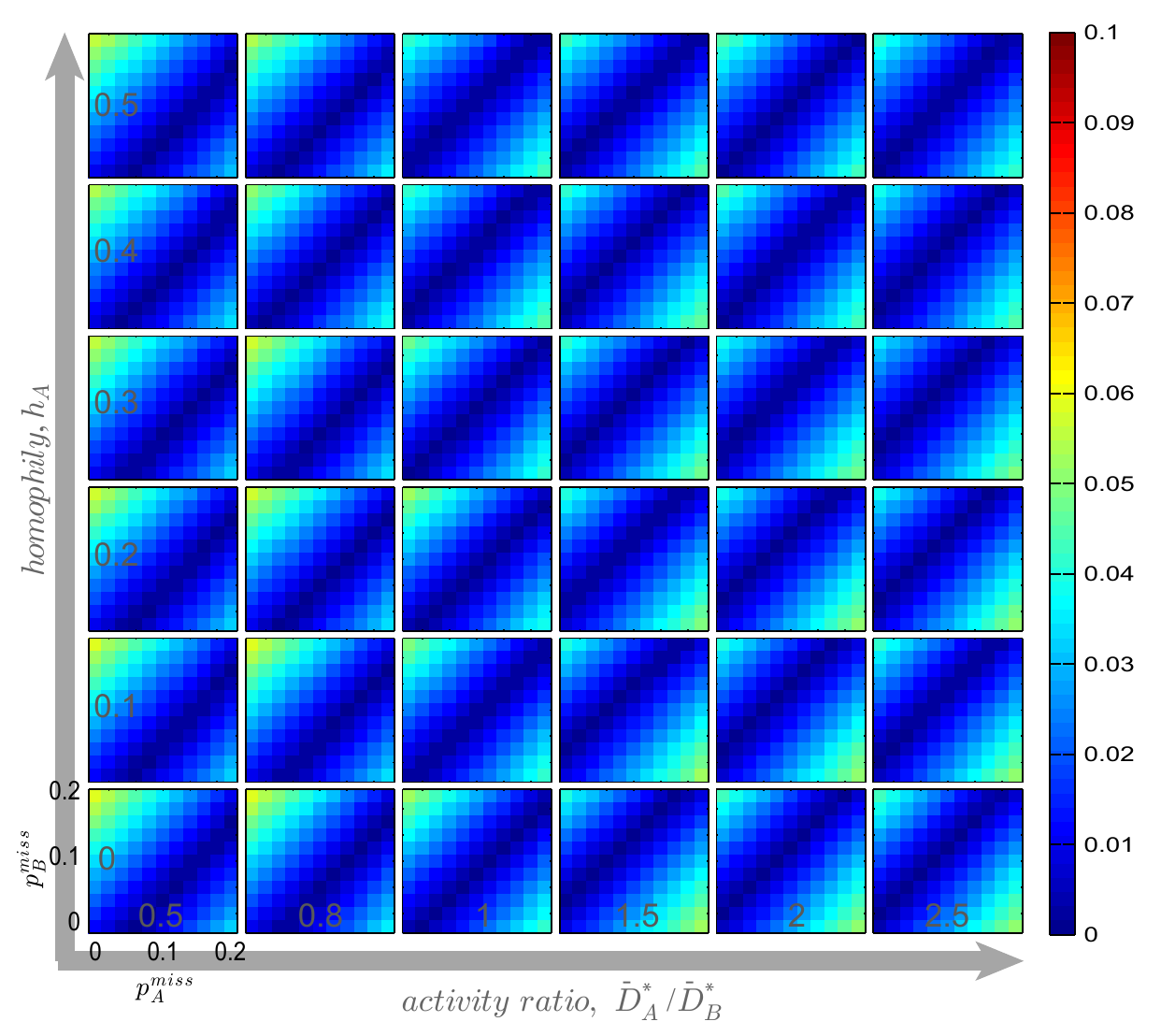}}
 \caption{Bias of $RDSI^{ego}$ on KOSKK network with random recruitment and \emph{degree reporting error}.}\label{Sfig7}
 \end{center}
 \end{figure}

 \clearpage

  \begin{figure}[h!]
 \begin{center}
 \centerline{\includegraphics[width=0.60\textwidth]{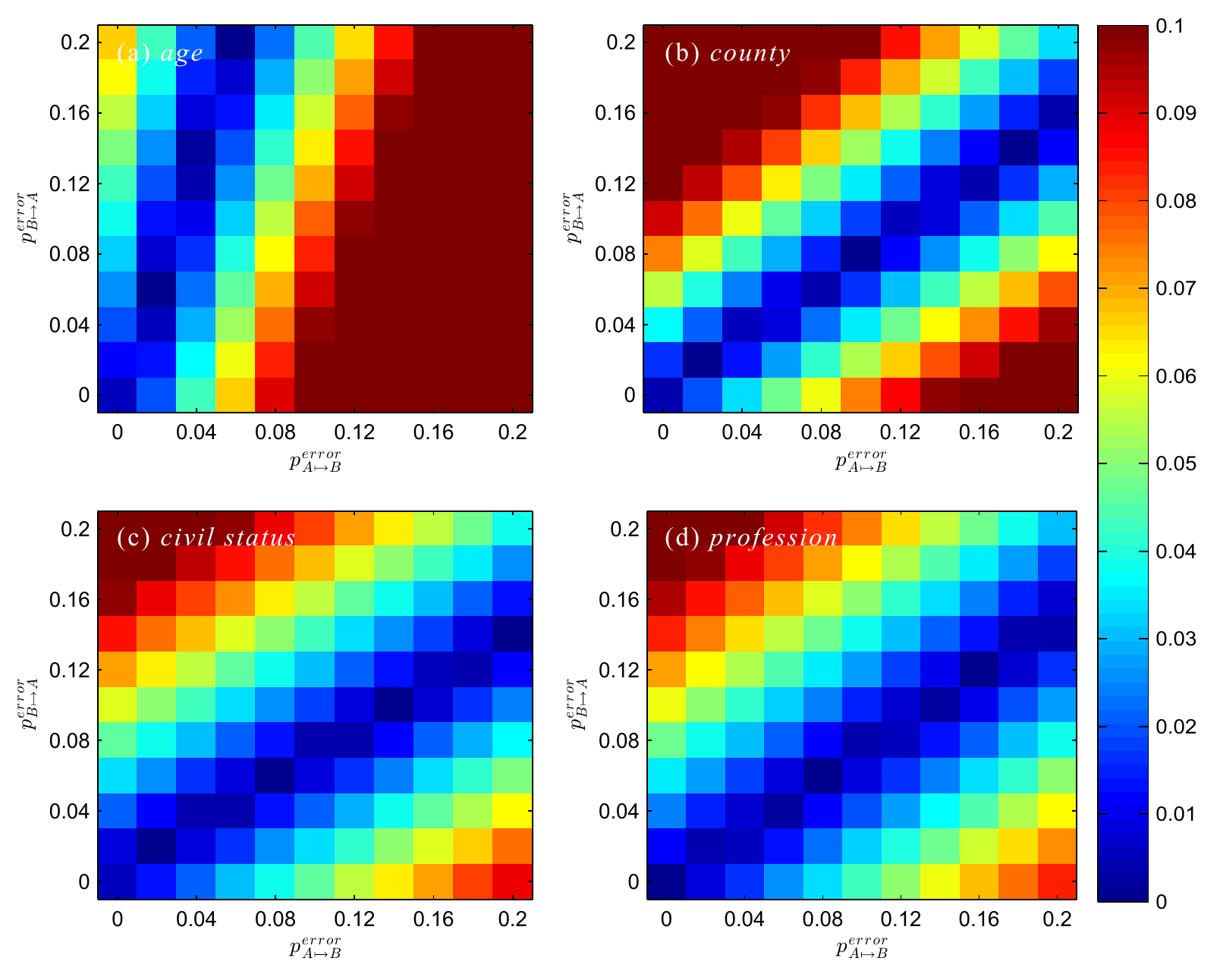}}
 \caption{Bias of $RDSI^{ego}$ on the MSM network with random recruitment and \emph{ego network reporting error}.}\label{Sfig8}
 \end{center}
 \end{figure}

  \begin{figure}[h!]
 \begin{center}
 \centerline{\includegraphics[width=0.60\textwidth]{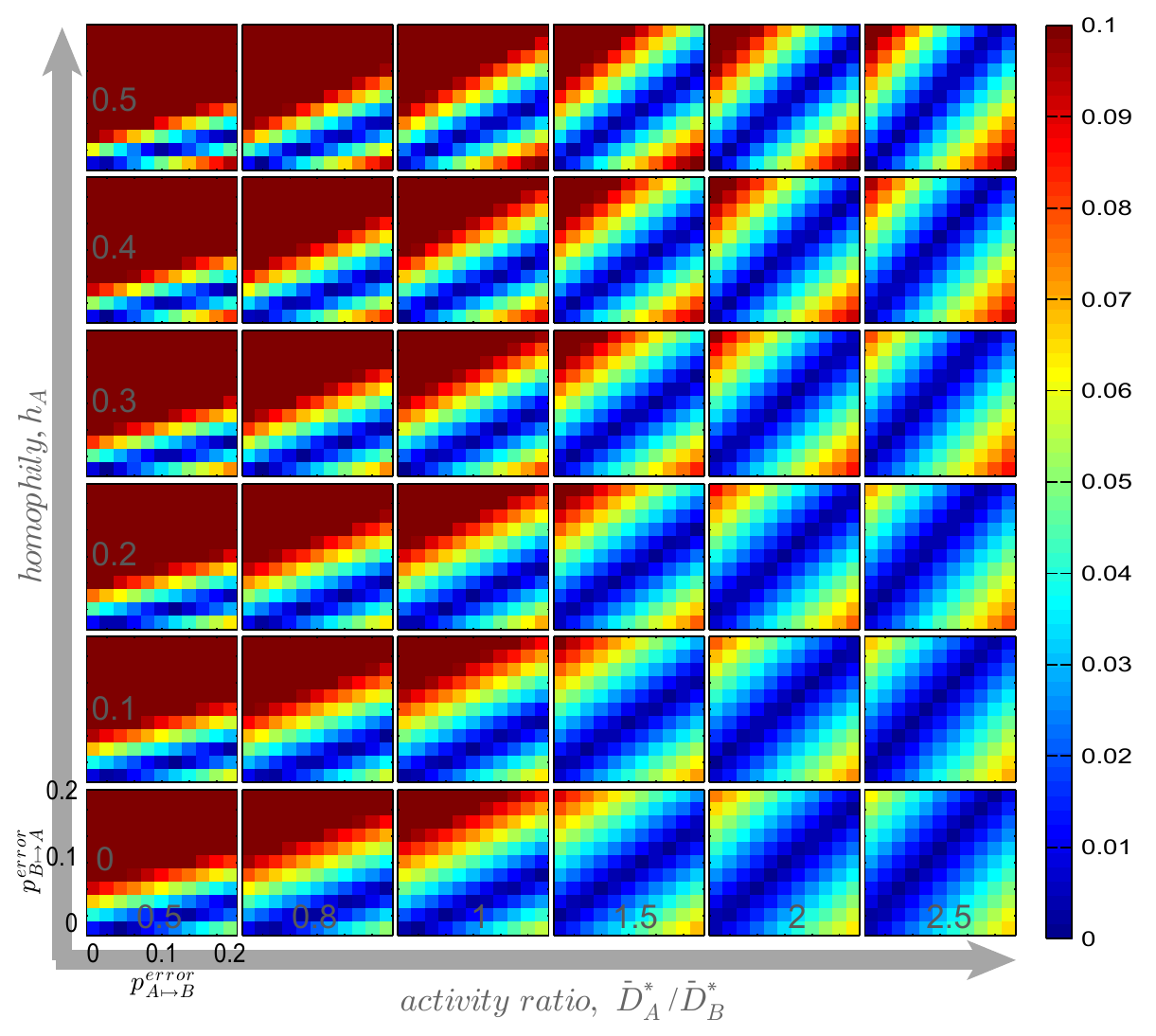}}
 \caption{Bias of $RDSI^{ego}$ on KOSKK network with random recruitment and \emph{ego network reporting error}.}\label{Sfig9}
 \end{center}
 \end{figure}

\clearpage
\bibliographystyle{vancouver}
\bibliography{RDS_EGO}

\end{document}